\documentclass[
reprint,
amsmath,amssymb,
aps,
prb,
longbibliography]{revtex4-2}

\usepackage{xspace}
\usepackage{graphicx}
\usepackage{bm}
\usepackage[colorlinks=true,urlcolor=blue,linkcolor=blue,citecolor=blue,bookmarks=false]{hyperref}
\usepackage{amsmath}
\usepackage{amssymb}
\usepackage{bbold}
\usepackage{soul}
\usepackage{dsfont}

\graphicspath{{./}}

\newcommand{\customref}[2]{\hyperref[#1]{\ref*{#1}#2}}

\usepackage{xcolor}

\definecolor{Ured}{HTML}{cc0000}
\definecolor{Ublue}{HTML}{1f65cf}
\definecolor{Ugreen}{HTML}{198a11}

\DeclareMathOperator{\Tr}{Tr}

\newcommand{\ie}[0]{i.e.\@\xspace}
\newcommand{\eg}[0]{e.g.\@\xspace}

\newcommand{\cf}[0]{cf.\@\xspace}

\newfont{\tensy}{cmsy10}

\renewcommand{\S}[0]{\hat{\mathcal{H}}}

\newcommand{\im}{\mathrm{i}}

\newcommand{\absolute}[1]{\left| #1 \right|}

\newcommand{\expvtext}[1]{\langle #1 \rangle}

\newcommand{\omegac}{\omega_\mathrm{c}}

\newcommand{\acri}{\alpha_\mathrm{CR1}}
\newcommand{\acrii}{\alpha_\mathrm{CR2}}
\newcommand{\aqci}{\alpha_\mathrm{QC1}}
\newcommand{\aqcii}{\alpha_\mathrm{QC2}}

\newcommand{\f}{\mathrm{F}}
\newcommand{\lo}{\mathrm{L}}
\newcommand{\lx}{\mathrm{L}_x}
\newcommand{\ly}{\mathrm{L}_y}
\newcommand{\lz}{\mathrm{L}_z}
\newcommand{\li}{\mathrm{L1}}
\newcommand{\lii}{\mathrm{L2}}

\newcommand{\cri}{\mathrm{CR1}}
\newcommand{\crii}{\mathrm{CR2}}
\newcommand{\qci}{\mathrm{QC1}}
\newcommand{\qcii}{\mathrm{QC2}}

\newcommand{\si}{s^\ast_1}
\newcommand{\sii}{s^\ast_2}

\newcommand{\siMPS}{s^\ast_\mathrm{1,MPS}}

\newcommand{\aii}{\alpha^\ast_2}

\newcommand{\artanh}{\mathrm{artanh}}

\begin{document}

\title{Tunable quantum criticality and pseudocriticality across the fixed-point annihilation in the anisotropic spin-boson model}

\author{Manuel Weber}
\affiliation{Institut f\"ur Theoretische Physik and W\"urzburg-Dresden Cluster of Excellence ct.qmat, Technische Universit\"at Dresden, 01062 Dresden, Germany}

\date{\today}

\begin{abstract}
Spin-boson models are simple examples of quantum dissipative systems, but also serve as effective models in quantum magnetism and exhibit nontrivial quantum criticality. Recently, they have been established as a platform to study the nontrivial renormalization-group (RG) scenario of fixed-point annihilation, in which two intermediate-coupling RG fixed points collide and generate an extremely slow RG flow near the collision. For the Bose Kondo model, a single $S=1/2$ spin where each spin component couples to an independent bosonic bath with power-law spectrum $\propto \omega^s$ via dissipation strengths $\alpha_i$, $i\in\{x,y,z\}$, such phenomena occur sequentially for the U(1)-symmetric model at $\alpha_z=0$ and the SU(2)-symmetric case at $\alpha_z = \alpha_{xy}$, as the bath exponent $s<1$ is tuned. Here we use an exact wormhole quantum Monte Carlo method for retarded interactions to explore how this nontrivial fixed-point structure affects the phase diagram and phase transitions of the anisotropic spin-boson model. In particular, we show how fixed-point annihilation within a symmetry-enhanced critical manifold leads to a variety of anisotropy-driven critical phenomena: (i) a continuous order-to-order transition beyond the Landau paradigm, (ii) a symmetry-enhanced first-order transition, and (iii) pseudocriticality. Depending on whether the attractive fixed point within the critical manifold corresponds to a critical or a localized phase, the transition between the two long-range ordered localized phases can be tuned from continuous to strongly first order and even becomes weakly first order in an extended regime close to the fixed-point collision. We extract critical exponents at the continuous transition, but also find scaling behavior at the symmetry-enhanced first-order transition, for which the inverse correlation-length exponent is given by the bath exponent $s$. Moreover, we provide direct numerical evidence for pseudocritical scaling on both sides of the fixed-point collision, which manifests in an extremely slow drift of the correlation-length exponent, even at the continuous transition. In addition, we also study the crossover behavior away from the SU(2)-symmetric case and determine the phase boundary of an extended U(1)-symmetric critical phase for $\alpha_z < \alpha_{xy}$. Our work establishes the spin-boson model as a paradigmatic example to access tunable criticality and pseudocriticality across the fixed-point collision in large-scale simulations, which is reminiscent of a scenario discussed in the context of deconfined criticality.
\end{abstract}

\maketitle

\section{Introduction}

Spin is one of the central properties of quantum matter and a building block for many applications in quantum magnetism, quantum optics, or quantum information. Since technological advances allow us to manipulate individual spins on the microscopic level, we need to understand how even a single spin
is affected by
the inevitable coupling to its environment. A simple class of Hamiltonians that capture the effects of quantum dissipation are spin-boson models \cite{RevModPhys.59.1}, in which a single spin is coupled to a  bath of harmonic oscillators with a gapless density of states $\propto \omega^s$.
The quantum dynamics of the spin can be tuned via the bath exponent $s$,
up to a point at which dissipation can induce nontrivial phases and quantum phase transitions in the spin degree of freedom \cite{doi:10.1080/14786430500070396}. Quantum dissipative impurity models also serve as effective models in various subdisciplines of condensed matter physics: they describe magnetic impurities in critical magnets \cite{Sachdev2479, PhysRevB.61.15152}, appear in the self-consistent solution of
extended dynamical mean-field theory which is used to explain
Kondo-breakdown transitions in heavy-fermion metals \cite{Si:2001aa, PhysRevB.68.115103},
and even exhibit connections to non-Fermi-liquid behavior in Sachdev-Ye-Kitaev models \cite{PhysRevLett.70.3339, RevModPhys.94.035004}. 
Recently, spin-boson models have been
identified as simple setups to study the nontrivial renormalization-group (RG) concept of fixed-point annihilation \cite{PhysRevLett.108.160401, PhysRevB.90.245130, PhysRevLett.130.186701, PhysRevB.106.L081109}.

Fixed-point annihilation is an RG scenario in which, as an external parameter $s$ is tuned that does not flow under the RG, two intermediate-coupling RG fixed points
collide and annihilate each other at $s=s^\ast$. 
It is a hallmark of this phenomenon that right after the collision
the RG flow is
exponentially suppressed 
in $\absolute{s-s^\ast}\ll 1$, 
providing a generic
mechanism to generate an extremely 
small order parameter
\cite{PhysRevD.80.125005}.
Fixed-point annihilation has been discussed
in various contexts in
high-energy physics \cite{PhysRevD.80.125005, PhysRevLett.32.292, Gies:2006aa, PhysRevD.90.036002, PhysRevD.94.094013, PhysRevD.94.025036, Giombi_2016, GUKOV2017583, Gorbenko:2018aa}, statistical mechanics \cite{PhysRevLett.43.737, PhysRevB.22.2560, PhysRevB.29.302, PhysRevB.69.134413, PhysRevLett.119.191602, 10.21468/SciPostPhys.5.5.050, PhysRevLett.131.131601, PhysRevE.108.064120, jacobsen2024lattice}, or condensed matter physics \cite{PhysRevLett.76.4588, PhysRevB.57.14254, PhysRevB.71.184519, PhysRevB.88.195119, PhysRevLett.113.106401, 2017PhRvB..95g5101J,
PhysRevX.5.041048, PhysRevX.7.031051, PhysRevB.102.201116, PhysRevB.102.020407, PhysRevB.97.214429, PhysRevB.100.134507, 2022arXiv220708744H, PhysRevB.107.165156, martin2023stable, doi:10.1126/sciadv.adr0634, hawashin2023nordicwalking},
but exact analytical or numerical studies in strongly interacting systems are rare.

Throughout the last years, fixed-point annihilation has gained interest in the study of an exotic type of quantum criticality that does not follow the Landau-Ginzburg-Wilson paradigm of conventional phase transitions.
In Landau theory,
a direct phase transition between two ordered states with distinct broken symmetries is generically first order and can only be continuous if the system parameters are fine tuned. However, it was proposed that fractionalized excitations can drive such a continuous non-Landau order-to-order transition in two-dimensional quantum magnets, a mechanism that is known as \textit{deconfined quantum criticality} \cite{Senthil1490, PhysRevB.70.144407, senthil2023deconfined}.
Early numerical studies of two-dimensional spin-$1/2$ model Hamiltonians found evidence
for a continuous phase transition between antiferromagnetic and valence-bond-solid order \cite{PhysRevLett.98.227202, PhysRevLett.100.017203},
but subsequent studies discovered unconventional scaling corrections \cite{PhysRevLett.104.177201, PhysRevB.88.220408, Jiang_2008, PhysRevLett.110.185701, PhysRevLett.101.050405, doi:10.1126/science.aad5007, PhysRevX.5.041048} that presumably hint towards a very weak first-order transition. Another feature of the deconfined quantum phase transition is an emergent SO(5) symmetry in the order-parameter fluctuations at criticality \cite{PhysRevLett.115.267203}, but emergent symmetries have also been found in related models with weak first-order transitions \cite{Zhao:2019aa, PhysRevB.99.195110}.
Fixed-point annihilation has been suggested
as one possible mechanism that explains the slow drift of critical exponents and an extremely small order parameter at the putative first-order transition in terms of the pseudocritical scaling experienced after the collision
\cite{PhysRevX.5.041048, PhysRevX.7.031051, PhysRevB.102.201116, PhysRevB.102.020407}; in this context, 
the external tuning parameter $s$ is the spatial dimension and the fixed-point collision is supposed to occur in a Wess-Zumino-Witten model tuned close to the two-dimensional case \cite{footnoteXXZspinboson}.
Numerical simulations of such systems face the challenge that their computational cost becomes increasingly expensive the higher the dimension and that it is hard to track the fixed-point collision if the external parameter only takes integer values. 
It has been pointed out that (0+1)-dimensional spin-boson models belong to the same hierarchy of Wess-Zumino-Witten models
\cite{PhysRevB.102.201116, PhysRevB.106.L081109}, in which the fixed-point collision can be tracked numerically with high precision \cite{PhysRevLett.130.186701}, as the bath exponent $s$ can be tuned continuously.
This raises the question if aspects of the (2+1)-dimensional transition can be captured by the spin-boson model.

In the original formulation of the spin-boson model, the dissipative bosonic bath only couples to one spin component, along which the spin gets localized for bath exponents $s<1$ and establishes long-range order with a finite local moment. Then, it is necessary to apply a transverse field to induce a quantum phase transition towards a delocalized phase \cite{KEHREIN1996313, PhysRevLett.91.170601, PhysRevLett.94.070604, PhysRevLett.102.249904, PhysRevLett.102.030601, PhysRevLett.108.160401},
which falls into the same universality class as the thermal phase transition in the one-dimensional Ising model with $1/r^{1+s}$ interactions \cite{PhysRevLett.29.917}. 
If multiple dissipation channels compete, 
frustration of different decoherence channels \cite{PhysRevLett.91.096401, PhysRevB.72.014417} can induce a plethora of novel phases and phase transitions, even in the absence of an external field. In particular, if two or three dissipation strengths are equal,
the spin can exhibit stable critical phases at weak coupling before it gets localized again at strong coupling. For each of these cases, the existence of a stable critical fixed point is an analytical prediction of the weak-coupling perturbative RG \cite{Sachdev2479, PhysRevB.61.15152, Smith_1999, PhysRevB.61.4041, PhysRevB.66.024426, PhysRevB.66.024427} and, as a result of nontrivial spin-Berry-phase effects, goes beyond the quantum-to-classical correspondence of the one-bath spin-boson model, whereas exact numerical techniques were required to find a localized phase beyond an unstable quantum-critical fixed point  \cite{PhysRevLett.108.160401, PhysRevB.90.245130, PhysRevB.87.125102, PhysRevB.100.014439, PhysRevLett.130.186701}. Altogether, one pair of intermediate-coupling
fixed points lies within the SU(2)-symmetric manifold of three identical dissipation strengths (this case is known as the Bose Kondo model), whereas another pair of fixed points lies within the U(1)-symmetric plane of the two-bath spin-boson model. As a function of the bath exponent $s$, each of these pairs exhibits an independent fixed-point annihilation \cite{PhysRevLett.108.160401, PhysRevB.90.245130, PhysRevLett.130.186701}. It is an open question as to how the two fixed-point collisions, which occur sequentially at distinct bath exponents $\si$ and $\sii$, affect the phase diagram and critical properties
of the anisotropic spin-boson model
away from the high-symmetry cases and across the symmetry-enhanced critical manifolds.

It is the purpose of this paper to close this gap. We use the recently developed wormhole quantum Monte Carlo (QMC) method \cite{PhysRevB.105.165129} to obtain exact numerical results for the anisotropic spin-boson model, in which two dissipation strengths are kept equal and the third coupling is tuned as an anisotropy. We map out the phase diagram along different cuts in parameter space and track the two sequential fixed-point annihilations, for which an approximate duality between weak- and strong-coupling fixed points enables us to determine the collision points with high precision. This allows us to establish a comprehensive RG picture for the anisotropic spin-boson model.
Among other things, we study the crossover from SU(2) to U(1) spins, for which a critical phase remains stable at finite spin anisotropies. 
\textit{Most importantly}, we explore the anisotropy-driven phase transitions across the high-symmetry critical manifold, in which the fixed-point annihilation occurs.
We find \textit{(i)} a \textit{symmetry-enhanced first-order transition} between two localized phases with distinct broken symmetries, \textit{(ii)} a \textit{non-Landau second-order quantum phase transition} between the same two ordered phases, and \textit{(iii)} \textit{pseudocritical scaling} in the vicinity of the fixed-point collision. The nature of these transitions is solely determined by the attractive fixed points within the critical manifold, for which anisotropy is always a relevant perturbation:
The symmetry-enhanced localized fixed point exhibits phase coexistence with a finite order parameter that combines the local moments of the distinct localized phases at finite anisotropy. The resulting first-order transition is described by a discontinuity fixed point that obeys finite-size scaling with an inverse correlation-length exponent given by the bath dimension $s$. By contrast, the high-symmetry critical fixed point has zero local moment and therefore turns this transition into a continuous one. 
Based on a finite-size-scaling analysis, we determine the in- and out-of-plane critical exponents at this fixed point to characterize the speed of the RG flow along different directions in parameter space.
In particular, close to the fixed-point collision the in-plane RG flow becomes so slow that we observe pseudocritical scaling in the anisotropic transition. We show that pseudocriticality occurs on both sides of the fixed-point collision, where the slow RG flow mimics scaling behavior with false critical exponents that only drift very slowly in our finite-size-scaling analysis. Our findings are consistent with a logarithmic drift of the pseudocritical exponents and a prefactor that depends linearly on the distance to the fixed-point collision.
All in all, fixed-point annihilation provides a generic mechanism to tune the nature of the same order-to-order transition from first- to second-order 
and to create a broad regime in which the transition becomes weakly first order without any need of fine tuning.
Our results reveal that already two competing dissipation channels are enough for a single spin to exhibit the full phenomenology of
strong or weak first- and second-order transitions at equal dissipation strength.

Our work establishes the anisotropic spin-boson model as an outstanding example in which the consequences of 
fixed-point annihilation on quantum criticality can be studied exactly using large-scale numerical simulations: its low dimensionality and continuous tunability via the bath exponent $s$ significantly enhance accessibility in simulations compared to many other model Hamiltonians.
Fixed-point annihilation within a high-symmetry critical manifold 
is
a generic and simple mechanism that enables but also intertwines a variety of exotic critical phenomena like symmetry-enhanced first-order transitions, continuous order-to-order quantum phase transitions beyond the Landau paradigm, and pseudocriticality leading to a weak first-order transition.
Thereby, the spin-boson model exhibits much of the phenomenology that has been discussed for two-dimensional quantum magnets in the context of deconfined criticality.
Although in our model the symmetry enhancement at criticality is not emergent but built in, all the characteristic features that are based on the fixed-point annihilation
remain universally applicable.
Moreover,
a single spin coupled to its environment is one of the simplest spin systems one can imagine 
and its nontrivial criticality adds complexity to the zoo of
fundamental spin models with nontrivial Berry-phase effects.
It is natural to ask how much of its physics can be found in more complicated setups, for which the spin-boson model provides an effective description. It also serves as a building block in higher-dimensional open quantum systems \cite{PhysRevLett.129.056402, PhysRevB.106.L161103, martin2023stable}, in which the competition between long-range interactions and the spin Berry phase plays an important role.

\subsection{Overview of this work}

The paper is organized as follows: In Sec.~\ref{Sec:Model}, we define the anisotropic spin-boson model and discuss our QMC approach. In Sec.~\ref{Sec:FPstructure}, we give a comprehensive overview of the fixed-point structure of the spin-boson model. To this end, we review previous analytical and numerical results and fill existing gaps using our QMC method.
In particular, we provide high-precision QMC results for the two sequential fixed-point collisions and the approximate fixed-point dualities, discuss limiting cases of the exact beta function, and summarize the complete RG picture in Sec.~\ref{Sec:RG_summary} in terms of schematic RG flow diagrams, which eventually determine the phases and phase transitions studied in this work.
In Sec.~\ref{Sec:PhaseDiagAni}, we use our QMC method to calculate the phase diagrams
along different cuts in parameter space,
but also
study the crossover behavior away from the SU(2)-symmetric parameter manifold towards the fixed points of the U(1)-symmetric model. 
In Sec.~\ref{Sec:QC_anisotropy}, we explore the anisotropy-driven phase transitions across the high-symmetry manifold and discuss the role of the fixed-point annihilation in tuning the order-to-order transition from second to first order via a broad regime in which the transition becomes weakly first order and exhibits pseudocritical scaling. We perform a finite-size-scaling analysis to calculate critical and pseudocritical exponents, but also give analytical arguments for the slow RG flow close to the fixed-point collision.
In Sec.~\ref{Sec:Ani}, we discuss how the fixed-point annihilation in the U(1)-symmetric model determines the phase transitions in the fully anisotropic spin-boson model. In Sec.~\ref{Sec:Con}, we summarize our findings, discuss relations to deconfined criticality in two-dimensional quantum magnets as well as to Bose-Fermi Kondo models, and conclude.
In Appendices \ref{App:RGquad} and \ref{App:pRG}, we provide additional analytical results for the fixed-point annihilation.

\section{Model and method\label{Sec:Model}}

We consider the anisotropic spin-boson model
\begin{align}
\label{eq:h}
\hat{H}
=
   \sum_{i = x,y,z} \sum_{q} \left[
\lambda_{q i} \hat{S}^i
( \hat{B}_{q i} + \hat{B}_{q i}^{\dagger} )
+
\omega_q \hat{B}_{q i}^{\dagger} \hat{B}_{q i}
\right] \, ,
\end{align}
where each of the three components $\hat{S}^i$ of a single $S=1/2$ spin
is coupled to an independent bosonic bath. Every bath is described by
an infinite number of harmonic oscillators, for which $\hat{B}_{q i}^\dagger$
($\hat{B}_{q i}$) creates (annihilates) a boson with frequency $\omega_q$ in the bath component $i$.
The bath spectra
$J_i(\omega) = \pi \sum_{q} \lambda_{q i}^{2} \delta(\omega -\omega_q)$
are of power-law form (here we take the continuum limit)
\begin{align}
\label{eq:spec}
J_i(\omega) = 2\pi\, \alpha_i \, \omega_\mathrm{c}^{1-s} \, \omega^s \,, \qquad
 0<\omega<\omega_\mathrm{c} =1 \, , \quad
 \end{align}
and the cutoff frequency $\omega_\mathrm{c}$ is taken as the unit of energy; 
beyond $\omega_\mathrm{c}$, we set $J_i(\omega)=0$.
The dimensionless
coupling parameters $\alpha_i$ measure the dissipation strength and, in the fully anisotropic case,
each component can take a different value. Throughout most parts of this paper, we consider the
U(1)-symmetric case with $\alpha_x=\alpha_y\equiv \alpha_{xy}$ and tune the spin anisotropy $\alpha_z$. At $\alpha_z=\alpha_{xy}$, the model becomes SU(2) symmetric.  We also define the anisotropy parameter $\Delta=1-\alpha_z/\alpha_{xy}$, where $\Delta=0$ corresponds to the isotropic case and $\Delta=1$ to $\alpha_z=0$.

For our simulations, we used an exact QMC method for retarded interactions \cite{PhysRevLett.119.097401, PhysRevB.105.165129}, 
which makes use of the fact that the bosonic baths can be traced out analytically. Our QMC method samples
a diagrammatic expansion of the partition function $Z = Z_\mathrm{b} \Tr_\mathrm{s} \hat{\mathcal{T}}_\tau \exp({-\S})$
in the spin degrees of freedom, where $Z_\mathrm{b}$ is the free-boson partition function and $\hat{\mathcal{T}}_\tau$ the time-ordering operator (note that we use the interaction representation \cite{PhysRevB.105.165129}). The retarded interaction vertex
\begin{align}
\label{Eq:Sret}
\S = - \iint_0^\beta d\tau d\tau' \sum_{i} K_i(\tau-\tau') \, \hat{S}^i(\tau) \, \hat{S}^i(\tau')
\end{align}
is nonlocal in imaginary time $\tau$ and mediated by the bath propagator
\begin{align}
K_i(\tau) =  \int_0^{\omega_\mathrm{c}} d\omega \frac{J_i(\omega)}{\pi} \frac{\cosh[\omega(\beta/2-\tau)]}{2 \sinh[\omega\beta/2]}
\end{align}
which fulfills $K_i(\tau+\beta)=K_i(\tau)$; here $\beta=1/T$ is the inverse temperature. For the power-law spectrum in Eq.~\eqref{eq:spec}, the bath propagator decays as $K_i(\tau) \propto 1/\tau^{1+s}$ for $\omegac \tau \gg 1$.
Our QMC method falls into the class of continuous-time methods \cite{RevModPhys.83.349} and our diagrammatic expansion has similarities to the hybridization-expansion method \cite{PhysRevLett.97.076405, PhysRevB.87.125102, 10.1063/1.4974328}.
However, the sampling of our expansion is based on the methodology developed for the stochastic series expansion \cite{PhysRevB.43.5950, PhysRevB.59.R14157, PhysRevE.66.046701}, which has been generalized to include imaginary times and retarded interactions \cite{PhysRevLett.119.097401}. During the diagonal updates, we use a Metropolis scheme to add or remove diagonal vertices $\hat{S}^z(\tau) \hat{S}^z(\tau')$ to or from the world-line configuration; the interaction range of $K_i(\tau-\tau')$ can be taken into account efficiently using inverse transform sampling \cite{PhysRevB.105.165129}. To implement the global directed-loop updates, we use the novel wormhole moves which transform the retarded diagonal vertices into spin-flip vertices (and vice versa) and allow for nonlocal tunneling of the loop head through a world-line configuration. For further details, we refer to Ref.~\cite{PhysRevB.105.165129} where the wormhole QMC method has been described comprehensively for the anisotropic spin-boson model. We also want to note that retarded spin interactions are often derived using a coherent-state representation; then, the resulting action includes an additional Berry-phase term, which is important to realize the critical fixed points described below \cite{PhysRevB.61.4041}. Our diagrammatic expansion in the interaction representation automatically takes this into account.

For a single spin degree of freedom, observables can only be accessed from imaginary-time correlation functions like
$\chi_i(\tau) = \langle \hat{S}^i(\tau) \hat{S}^i(0) \rangle$.
From this, we calculate the dynamical spin susceptibility
\begin{align}
\chi_i(\im \Omega_n)
=
\int_0^\beta d\tau \, e^{\im \Omega_n \tau}
\expvtext{\hat{S}^i(\tau) \hat{S}^i(0)}
\end{align}
directly in Matsubara frequencies $\Omega_n = 2\pi n / \beta$, $n\in \mathds{Z}$.
For $i\in\{x,y\}$, the susceptibilities can be calculated during the propagation of the directed loop, whereas the $z$ component is determined from the world-line configuration.
We also define the static susceptibility $\chi_i = \chi_i(\im\Omega_0)$ which can be used to identify the formation of a local moment. Then, $\chi_i(T) = m_i^2/T$ approaches a Curie law for temperatures $T\to 0$. A finite-temperature estimate of the local moment can also be obtained from 
\begin{align}
\label{eq:locmom}
m^2_i(T)
=
\expvtext{\hat{S}^i(\beta/2) \hat{S}^i(0)} \, ,
\end{align}
which needs to be extrapolated towards zero temperature and is indicative of long-range order in the imaginary-time direction.

To study the critical properties of the spin-boson model, it is also useful
to calculate the correlation length along imaginary time (correlation time)
\begin{align}
\label{eq:clength}
\xi_i
=
\frac{1}{\Omega_1} \sqrt{\frac{\chi_i(\im \Omega_0)}{\chi_i(\im\Omega_1)} - 1} \, .
\end{align}
It is defined in analogy to the spatial correlation length \cite{doi:10.1063/1.3518900}, because space and imaginary time can be treated on the same level for quantum problems. While the spatial correlation length is evaluated from the equal-time correlations in momentum space at the ordering vector $Q$ and the nearest vector shifted by the momentum resolution $\delta q = 2\pi / L$, the correlation time is calculated from the dynamical correlations in Matsubara space at the ordering component $\Omega_0=0$ and the nearest component $\Omega_1 = 2\pi / \beta$ which is shifted by the resolution of Matsubara frequencies. The system size along imaginary time is $\beta$, therefore, the renormalized correlation length $\xi_i / \beta$ diverges (scales to zero) in the ordered (disorderd) phase. At criticality, the correlation length becomes scale invariant and $\xi_i / \beta$ approaches a constant. A closely related measure is the correlation ratio
\begin{align}
\label{eq:cratio}
R_i
	=
	1 - \frac{\chi_i(\im \Omega_1)}{\chi_i(\im \Omega_0)} \, ,
\end{align}
which scales to one (zero) in the ordered (disordered) phase and becomes RG invariant at criticality.
 
\section{Fixed-point structure\label{Sec:FPstructure}}

The phase diagram and critical properties of the spin-boson model can be understood from the underlying
RG structure. Therefore, we first review what is known from analytical and numerical studies and complete the missing parts of this picture using our QMC method.

\subsection{Fully anisotropic spin-boson model\label{Sec:FPstructureAni}}

At zero dissipation, our system in Eq.~\eqref{eq:h} is described by the free-spin fixed point $\f$ located at $\vec{\alpha}_\f =(0,0,0)$,
where the static susceptibility $\chi_i(T) = m^2_i /T$ follows a Curie law with local moment $m^2_i = 1/4$ for all $i\in\{x,y,z\}$.
For bath exponents $s>1$, the coupling to the bath is an irrelevant RG perturbation, so that $\f$ remains stable for any dissipation strength $\alpha_i$; the main effect of the bath is to renormalize $m^2_i < 1/4$, which will be shown in Sec.~\ref{Sec:LocalMom}. For $s<1$, the coupling to the bath is a relevant perturbation, so that $\f$ is unstable for any $\alpha_i>0$. For the fully anisotropic case ($\alpha_x \neq \alpha_y \neq \alpha_z$), the system flows to one of the three stable strong-coupling fixed points $\lo_i$, $i\in\{x,y,z\}$, chosen according to the strongest dissipation strength $\alpha_i$ \cite{PhysRevB.66.024426, PhysRevB.66.024427}; 
each of these fixed points describes a localized phase with spontaneously broken $\mathds{Z}_2$ symmetry along spin orientation $i$. Again, the static susceptibility follows a Curie law with $m^2_i > 0$ but $m^2_{j\neq i} = 0$. 
In most parts of our paper, we will encounter only one localized phase along one of the three spin orientations $i\in\{x,y,z\}$; for simplicity, we will denote this fixed point by $\lo$.

\subsection{Stable intermediate-coupling fixed points within the high-symmetry manifolds}
\label{Sec:pertRG}

If at least two components of the dissipation strength $\alpha_i$ are equal, the fixed-point structure of the spin-boson
model becomes more complex, which was first studied using the weak-coupling perturbative RG
\cite{Smith_1999, PhysRevB.61.4041, PhysRevB.61.15152, PhysRevB.66.024426, PhysRevB.66.024427}.
Expanding about the marginal point at $\vec{\alpha}_\f =(0,0,0)$ and $s=1$,
the two-loop beta function for one of the coupling parameters becomes
\cite{PhysRevB.66.024426, PhysRevB.66.024427}
\begin{align}
\beta(\alpha_x) \equiv 
\frac{d\alpha_x}{d\ln \mu}
=
- \alpha_x \left[\left(1-s\right) - 2 \left(\alpha_y + \alpha_z\right) \left(1 - 2 \alpha_x\right) \right] \, ,
\label{eq:beta}
\end{align}
whereas 
$\beta(\alpha_y)$ and $\beta(\alpha_z)$
are obtained by cyclic permutation of the indices $i\in\{x,y,z\}$
\footnote{Our definition of the couplings $\alpha_i$ differs by a factor of two from Refs.~\cite{PhysRevB.66.024426, PhysRevB.66.024427}, so that Eqs.~\eqref{eq:cr1} and \eqref{eq:cr2} match our numerical results.}%
.
The beta function describes how the effective couplings $\alpha_i$
renormalize as the reference scale $\mu$ is changed under an RG step.
The zeros of the beta function correspond to fixed points under an RG transformation
and can describe stable phases or phase transitions.

For $s<1$, the coupled
flow equations contain three equivalent nontrivial fixed points $\cri$ within the three U(1)-symmetric planes, in which one coupling is zero.
One of them lies within the $xy$ plane at
\begin{align}
\nonumber
\vec{\alpha}_\cri &= (\acri, \acri, 0) \quad \mathrm{with} \\
\acri &= \frac{1-s}{2}+\frac{(1-s)^2}{4} + \mathcal{O}[(1-s)^3] \, ,
\label{eq:cr1}
\end{align}
whereas the other ones in the $yz$ and $xz$ planes can be obtained accordingly.
Without loss of generality, we restrict our discussion to the fixed point in the $xy$ plane.
$\cri$ is stable towards perturbations that conserve
the U(1) symmetry, \ie, in the $\alpha_z$ direction, but unstable towards
anisotropies in the $x$ and $y$ directions.

Moreover, there is an additional fixed point at
\begin{align}
\nonumber
\vec{\alpha}_\crii &= (\acrii, \acrii, \acrii) \quad \mathrm{with} \\
\acrii &= \frac{1-s}{4}+\frac{(1-s)^2}{8} + \mathcal{O}[(1-s)^3] \, ;
\label{eq:cr2}
\end{align}
$\crii$ is a stable fixed point within the SU(2)-symmetric parameter
manifold $(\alpha,\alpha,\alpha)$, but any perturbation that breaks
this symmetry will lead away from $\vec{\alpha}_\crii$.
To simplify our notation throughout this work, we will sometimes refer to $\cri$ and $\crii$ as stable fixed points, but always mean within their high-symmetry manifold.

The two fixed points
$\cri$ and $\crii$ describe critical phases in which the long-range
decay of the spin
autocorrelation function fulfills $\chi_i(\tau) 
\propto \tau^{- \eta_i}$.
Note that $\eta_i = 1-s$ is an exact result from
the diagrammatic structure of the susceptibility \cite{PhysRevB.61.15152, PhysRevB.66.024426}
that is valid for $i\in\{x,y\}$ at $\cri$ and $i\in\{x,y,z\}$ at $\crii$, whereas at $\cri$ the exponent
$\eta_z = 2(1-s)+ (1-s)^2 + \mathcal{O}[(1-s)^3]$ is only known perturbatively near $s=1$ \cite{PhysRevB.66.024427}. As a result, the static susceptibility
fulfils $\chi_i(T) \propto T^{-\tilde{\eta}_i}$ with $\tilde{\eta}_i = 1- \eta_i$, \ie, the critical phases have
a local moment of $m^2_i = 0$, and the low-frequency part of the dynamical
susceptibility becomes $\chi_i(\im\Omega_n) \propto \absolute{\im \Omega_n}^{-\tilde{\eta}_i}$.
Hence, $\chi_i(\im \Omega_1) \propto T^{-\tilde{\eta}_i}$, so that the normalized correlation length $\xi_i /\beta$ defined in Eq.~\eqref{eq:clength} is finite.

\subsection{Unstable intermediate-coupling fixed points, fixed-point annihilation, and duality}

Large-scale numerical studies revealed that the perturbative RG picture described above is not yet complete.
For the U(1)-symmetric spin-boson model at $\alpha_z=0$,
an MPS approach determined the phase diagram and found,
in addition to a critical phase described by $\cri$, a localized
phase $\li$ where the U(1) symmetry
is spontaneously broken \cite{PhysRevLett.108.160401, PhysRevB.90.245130}.
The fixed point $\li$ appears at infinite coupling and is separated from the critical
phase $\cri$ via an unstable quantum critical fixed point $\qci$.
Later, a strong-coupling localized phase $\lii$
was also identified in the SU(2)-symmetric model using QMC simulations \cite{PhysRevB.87.125102},
which again is separated from the critical phase $\crii$ via a quantum critical fixed point $\qcii$ \cite{ PhysRevB.100.014439, PhysRevLett.130.186701}. In the low-temperature limit, the localized phases $\li$ and $\lii$ follow a Curie law $\chi_i(T) = m^2_i / T$ with a finite local moment $m^2_i > 0$ along the symmetry-broken spin orientations, whereas $m^2_z = 0$ for $\li$. The low-frequency part of the dynamical spin susceptibility
$\chi_i(\im\Omega_n) \propto \absolute{\im \Omega_n}^{-s}$
still resembles the gapless features of the corresponding critical phase \cite{PhysRevB.87.125102,  PhysRevLett.130.186701}, but $\chi_z(\im\Omega_n)$ approaches a constant for $\li$.
The gapless excitations in $\chi_i(\omega)$ can be interpreted as the Goldstone modes that occur due to spontaneous symmetry breaking of the continuous rotational symmetry of spin plus bath; note that this signature is absent in the $\lo$ phase which only breaks a $\mathds{Z}_2$ symmetry.
Moreover, our estimator for the normalized correlation length
$\xi_i / \beta$ in Eq.~\eqref{eq:clength} diverges [because $\chi_i(\im \Omega_0)\propto T^{-1}$ and $\chi_i(\im \Omega_1) \propto T^{-s}$],
whereas $\xi_z / \beta \to 0$ for $\li$.

It was first suggested for the U(1)-symmetric model that the two intermediate-coupling
fixed points $\cri$ and $\qci$ approach each other, as the bath exponent $s$ is reduced,
and eventually collide and annihilate each other \cite{PhysRevLett.108.160401, PhysRevB.90.245130}.
While Refs.~\cite{PhysRevLett.108.160401, PhysRevB.90.245130} provided indirect evidence via the disappearance of $\qci$, the fixed-point collision has been tracked directly for the SU(2)-symmetric case \cite{PhysRevLett.130.186701}. Analytical confirmation of the fixed-point collision has also been obtained in a large-$S$ limit of the SU(2)-symmetric model \cite{PhysRevB.106.L081109}.
In the following, we will review previous results for the SU(2)-symmetric model and repeat our QMC simulations for the U(1)-symmetric case, to provide a complete picture for the fixed-point annihilation in the anisotropic spin-boson model.

\begin{figure}[t]
\includegraphics[width=\linewidth]{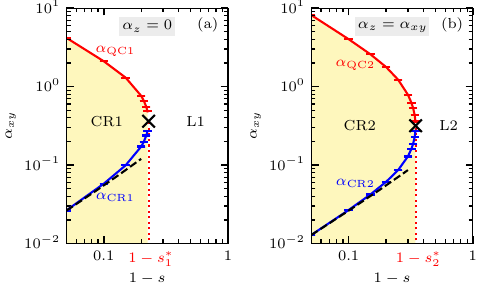}
\caption{%
Fixed-point structure at intermediate spin-boson couplings $\alpha_{xy}$ and for bath exponents $0<s<1$.
Nontrivial fixed points only occur for (a) $\alpha_z=0$ and (b) $\alpha_z=\alpha_{xy}$.
As a function of $s$, the two fixed points CR1/2 and QC1/2 collide and annihilate each other at
$s^\ast_{1/2}$; the precise values at which the collisions occur are determined in Fig.~\ref{fig:FPduality}.
The black dashed lines indicate the predictions \eqref{eq:cr1} and \eqref{eq:cr2} of the perturbative RG for $\acri$ and $\acrii$.
Results for the SU(2)-symmetric case presented in (b) are taken from Ref.~\cite{PhysRevLett.130.186701}.
}
\label{fig:FPannihilation}
\end{figure}

The evolution of the two intermediate-coupling fixed points as a function of the bath exponent $s$ is summarized
in Fig.~\ref{fig:FPannihilation}(a)
for $\alpha_z=0$ and in Fig.~\ref{fig:FPannihilation}(b) for $\alpha_z=\alpha_{xy}$.
Results have been obtained from a finite-size-scaling analysis of the spin susceptibility,
as described in detail in Ref.~\cite{PhysRevLett.130.186701} and its Supplemental Material
(we will also discuss in Sec.~\ref{Sec:phasediagrams} how the fixed points become accessible via the spin susceptibility).
At small $\alpha$ and $1-s$, the evolution of $\acri$ and $\acrii$ in Fig.~\ref{fig:FPannihilation} agrees well with
the predictions \eqref{eq:cr1} and \eqref{eq:cr2} of the perturbative RG, whereas at
larger couplings they start to deviate. It is apparent that the fixed-point collision
takes place at different $s^\ast_1$ and $s^\ast_2$ for the two cases, which will have important consequences for the phase diagram and critical properties of the anisotropic spin-boson model.

\begin{figure}[t]
\includegraphics[width=\linewidth]{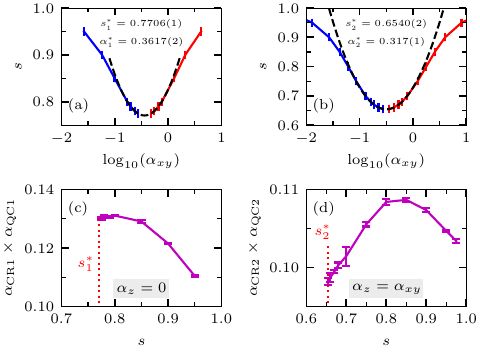}
\caption{%
(a), (b) Estimation of the coordinates $(s^\ast_{1/2},\alpha^\ast_{1/2})$, at which the intermediate-coupling fixed points collide,
by fitting their evolution to
$s(\alpha_{xy}) = s^\ast + (b/a) \ln^2(\alpha_{xy} / \alpha^\ast)$.
Final estimates for $s^\ast_{1/2}$ and $\alpha^\ast_{1/2}$ are stated in  (a) and (b),
whereas $b_1 / a_1 = 0.0532(2)$
and $b_2 / a_2 = 0.0562(5)$.
(c),(d) Numerical test of the fixed-point duality. If the duality was exact, the product $\alpha_\mathrm{CR1/2}\times\alpha_\mathrm{QC1/2}$ would be constant.
Results for the SU(2)-symmetric case presented in (b) and (d) are taken from Ref.~\cite{PhysRevLett.130.186701}.
}
\label{fig:FPduality}
\end{figure}

To determine the precise coordinates ($s^\ast_{1/2},\alpha^\ast_{1/2}$) of the fixed-point collisions, we make use of the approximate
symmetry of the fixed-point evolution on the logarithmic $\alpha_{xy}$ scale, which is apparent
from Fig.~\ref{fig:FPannihilation} and has first been
observed for the SU(2)-symmetric model \cite{PhysRevLett.130.186701}.
Figures \ref{fig:FPduality}(a) and \ref{fig:FPduality}(b)
show quadratic fits close to the fixed-point collision.
In this way, $\sii=0.6540(2)$ has been obtained in Ref.~\cite{PhysRevLett.130.186701}
and now we determine $\si=0.7706(1)$. Our estimate
for $\si$ improves the previous MPS result $\siMPS = 0.76(1)$ \cite{PhysRevLett.108.160401, PhysRevB.90.245130} because we are able to extract $\si$ from fitting the functional form of the fixed-point collision (for details see Ref.~\cite{PhysRevLett.130.186701}).

Our numerical data reveal an approximate duality between the weak- and strong-coupling fixed points. If the duality was exact, the product $\alpha_\mathrm{CR1/2}\times\alpha_\mathrm{QC1/2}$ would be a constant that is independent of the bath exponent $s$. Figures \ref{fig:FPduality}(c) and \ref{fig:FPduality}(d) show this product for the two fixed-point collisions. Given the fact that the individual fixed-point couplings vary by several orders of magnitude, their product is almost constant. In particular, the U(1)-symmetric case depicted in Fig.~\ref{fig:FPduality}(c) only shows little deviations near the fixed-point collision. It was conjectured that this duality is a symmetry of the beta function \cite{PhysRevLett.130.186701} and therefore allows for the prediction of critical exponents at the strong-coupling fixed point based on perturbative results at the weak-coupling fixed point. Moreover, Ref.~\cite{PhysRevLett.130.186701} identified that this duality becomes exact in the limit of large total spin $S\to\infty$, as apparent from the analytical beta function of Refs.~\cite{Cuomo:2022aa, Beccaria_2022, PhysRevB.106.L081109}.
Duality relations often appear in classical and quantum spin systems \cite{NishimoriOrtiz}
and had been identified for a single quantum rotor coupled to a dissipative bath \cite{PhysRevLett.51.1506, Falci:1999aa}. While dualities often pinpoint the phase transition to appear at the self-dual point \cite{NishimoriOrtiz}, the spin-boson model seems to exhibit an approximate duality between two fixed points.

In close vicinity to the fixed-point collision, the exact (but unknown) beta function within the high-symmetry manifold can always be expanded up to quadratic order in the coupling $\bar{\alpha}$, such that
\begin{align}
\label{eq:RGquadratic}
\beta(\bar{\alpha}) \equiv \frac{d\bar{\alpha}}{d \ln \mu} = 
 a \left(s - s^\ast \right) - b \left( \bar{\alpha} - \bar{\alpha}^\ast \right)^2 \, ,
\end{align}
where $a$ and $b$ are expansion coefficients. In general, $\bar\alpha$ is the coupling
constant of our system, \ie, $\bar\alpha = \alpha_{xy}$ for the U(1)-symmetric model (at $\alpha_z=0$)
and $\bar\alpha = \alpha_{xy} = \alpha_z$ for the SU(2)-symmetric case. 
Our fixed-point duality even suggests $\bar{\alpha} = \ln \alpha_{xy}$, so that Eq.~\eqref{eq:RGquadratic}
is valid in a rather broad regime of bath exponents $s$.
The beta function in Eq.~\eqref{eq:RGquadratic} is just a parabola opened downwards, which can be shifted up and down using the bath exponent $s$.
For $s\approx s^\ast$, $\beta(\bar\alpha^\ast) \approx 0$ and $\beta^\prime(\bar\alpha^\ast) = 0$,
leading to an extremely slow RG flow near $\bar{\alpha} = \bar{\alpha}^\ast$. A detailed solution of Eq.~\eqref{eq:RGquadratic} and its characteristic RG flow is given in Appendix~\ref{App:RGquad}. In particular, Eq.~\eqref{eq:RGquadratic} justifies the fitting form for the fixed-point evolution used in Fig.~\ref{fig:FPduality} and predicts $1/\nu \propto (s-s^\ast)^{1/2}$ for the inverse correlation-length exponents at the two intermediate-coupling fixed points. 
We will make use of Eq.~\eqref{eq:RGquadratic} in Sec.~\ref{Sec:Pseudocriticality}, when we discuss the characteristic RG flow close to the fixed-point collision.

At this point, it is also worth mentioning that the fixed-point collision is already contained in the two-loop beta function of Eq.~\eqref{eq:beta}, although the perturbative RG has no predictive power for $s\ll 1$ or for the strong-coupling fixed points $\qci$ and $\qcii$ \cite{PhysRevB.90.245130, PhysRevLett.130.186701}. Nevertheless, some of the qualitative aspects near the fixed-point collision are captured correctly. The solution of Eq.~\eqref{eq:beta} is summarized in Appendix~\ref{App:pRG} and we will come back to this when discussing the RG flow close to the fixed-point annihilation in Sec.~\ref{Sec:Pseudocriticality}.

\subsection{Renormalization-group flow diagrams and their consequences for criticality}
\label{Sec:RG_summary}

\begin{figure*}[t]
\includegraphics[width=\linewidth]{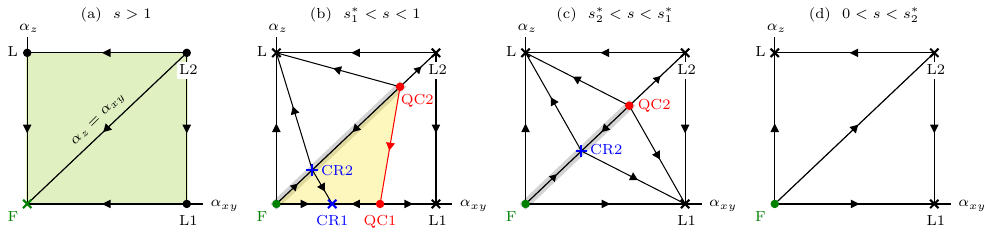}
\caption{%
Schematic illustration of the RG flow for the anisotropic spin-boson model as a function of the dissipation strengths $\alpha_{xy}$ and $\alpha_z$ within the different regimes tuned by the bath exponent $s$.
Stable (unstable) fixed points (within their symmetry sector) are marked by crosses (circles) and the direction of the RG flow is indicated by the arrowheads.
(a) For $s>1$, the free-spin fixed point $\f$ is stable for all $\alpha_i$, as indicated by the green shaded area.
(b) For $\si < s < 1$, $\f$ becomes unstable. There appear two pairs of intermediate-coupling fixed points, one pair on the diagonal line ($\alpha_z = \alpha_{xy}$) and the other at $\alpha_z = 0$, each consisting of a critical phase CR1/2 and a quantum critical point QC1/2. Note that $\crii$ and $\qcii$ are unstable towards perturbations that break the SU(2) spin symmetry and lead to the stable fixed points of the localized phases $\lo$ and $\li$ as well as to the stable critical phase $\cri$. The yellow (gray) shaded area indicates the extent of the critical phase $\cri$ ($\crii$) and the red line the separatrix between the $\cri$ and $\li$ phases.
(c) For $\sii < s < \si$, the fixed points $\cri$ and $\qci$ have disappeared via fixed-point annihilation. Only $\crii$ and $\qcii$ remain, which are still unstable towards symmetry-breaking perturbations leading to $\lo$ and $\li$.
(d) For $0<s < \sii$, also the second pair of fixed points has disappeared, so that the localized phases $\lo$, $\li$, and $\lii$ are the only stable phases within their symmetry sectors.
}
\label{fig:RGflow}
\end{figure*}

The final fixed-point structure of the anisotropic spin-boson model including the RG flow between each of the fixed points is summarized in Fig.~\ref{fig:RGflow},
for which a detailed description is given in the caption. In the following,
we want to discuss the consequences for the phase diagram and critical properties.

The spin-boson model exhibits four different regimes that can be distinguished by the bath exponent $s$.
For the superohmic regime with bath exponent $s>1$, the RG flow depicted in Fig.~\ref{fig:RGflow}(a)
illustrates that the free-spin fixed point $\f$ remains stable for all dissipation strengths. Consequently, the spin anisotropy $\alpha_i$ cannot drive any quantum phase transitions for $s>1$.

In the subohmic regime $0<s<1$, the RG flow is determined by the two fixed-point collisions that occur sequentially at $\si$ and $\sii$.
For $\si < s < 1$, the nontrivial RG flow, which includes two pairs of intermediate-coupling fixed points, is depicted in Fig.~\ref{fig:RGflow}(b). For $\alpha_{z} < \alpha_{xy}$, there exist two stable phases, $\cri$ and $\li$, separated by a line (red) at which a quantum phase transition occurs whose critical properties are solely characterized by the unstable fixed point $\qci$. In particular, universal properties like critical exponents have to be the same for all anisotropies crossing the separatrix.
The criticality of the U(1)-symmetric model has been studied in great detail at $\alpha_z = 0$ \cite{PhysRevB.90.245130} and will not be repeated in this work.
It is the purpose of Sec.~\ref{Sec:PhaseDiagAni} to determine the precise phase boundary of the $\cri$--$\li$ quantum phase transition at finite anisotropies and to confirm the universality of the critical exponent along this line. Moreover, we will study the crossover between the SU(2)-symmetric fixed points and the U(1)-symmetric ones.

Beyond the first fixed-point collision, \ie, for $\sii < s < \si$ illustrated in Fig.~\ref{fig:RGflow}(c), the localized fixed points $\lo$ and $\li$ describe the only stable phases for $\alpha_z \neq \alpha_{xy}$. Only within the SU(2)-symmetric manifold we will find nontrivial fixed points. The critical properties of the fully symmetric model have been studied in detail in Ref.~\cite{PhysRevLett.130.186701}.
Eventually, for $0<s<\sii$ the last pair of intermediate-coupling fixed points has disappeared due to the second fixed-point collision, so that even the critical behavior at $\alpha_{z} = \alpha_{xy}$ is determined by a localized fixed point.

The RG flow diagrams in Fig.~\ref{fig:RGflow} also determine the nature of the quantum phase transitions
driven by the spin anisotropy $\alpha_z / \alpha_{xy}$ through the SU(2)-symmetric critical manifold.
For $0<s<\sii$, we always find an $\li$--$\lo$ transition between two ordered phases
that are separated by a first-order transition because at the high-symmetry point the system
is in the $\lii$ phase where both order parameters $m^2_{xy}$ and $m^2_z$ are equal and therefore coexist. This symmetry-enhanced first-order transition is described by a discontinuity fixed point, for which we expect to find finite-size scaling relations.
Such a first-order transition also occurs for $\sii < s < \si$ and $\si < s < 1$ in regimes where the high-symmetry fixed point is $\lii$. However, for $\sii < s < \si$ depicted in Fig.~\ref{fig:RGflow}(c) there exists a regime in which the two localized phases are separated by the critical fixed point $\crii$ with $m^2_{xy}=m^2_{z}=0$. As a result, the two ordered phases are separated by a continuous transition.
Furthermore, for $\si < s < 1$ in Fig.~\ref{fig:RGflow}(b) we find a continuous transition between the critical phase $\cri$ and the localized phase $\lo$. We will study these anisotropy-driven quantum phase transitions in Sec.~\ref{Sec:QC_anisotropy}. In particular, we find that the fixed-point annihilation provides us with a tunable first-order to continuous $\li$--$\lo$ transition that can exhibit an arbitrarily weak first-order regime close to the fixed-point annihilation without fine tuning. In this regime, we will also be able to study the pseudocritical RG scaling in detail.

\subsection{Critical exponents and finite-size scaling}

The quantum phase transitions of the anisotropic spin-boson model and their critical properties are determined by the fixed points $\qci$, $\crii$, and $\lii$
which have one irrelevant and one relevant RG direction [within the U(1)-symmetric parameter space], as indicated by the ingoing and outgoing arrowheads in Fig.~\ref{fig:RGflow}, respectively. 
In close vicinity to these fixed points, the RG equations can be linearized and decoupled so that the RG flow along the (ir)relevant scaling variables becomes $\Delta\alpha_{i}(\mu) \propto \mu^{-1/\nu_i}$ with $i\in\{\parallel, \perp\}$. Here, $\Delta\alpha_\parallel$ denotes the iterated distance along the high-symmetry direction as the RG scale $\mu$ is reduced,
whereas $\Delta\alpha_\perp$ describes the perpendicular flow. Speed and direction of the RG flow is determined by the inverse correlation-length exponents $1/\nu_i$; a positive (negative) exponent describes a relevant (irrelevant) perturbation. All points in parameter space that flow into the fixed point along the direction for which $1/\nu_i < 0$ belong to the critical manifold of this fixed point; in Fig.~\ref{fig:RGflow} these are, \eg, the red line for $\qci$ or the gray shaded line for $\crii$. Then, all paths that cross the critical manifold will eventually converge to the flow line determined by the relevant direction at the fixed point so that the corresponding $1/\nu_i > 0$ dictates the scaling at the phase transition. For the fixed points 
$\qci$, $\crii$, and $\lii$ we denote the critical exponents by $1/\nu_{\qci \parallel}$, $1/\nu_{\crii \perp}$, and $1/\nu_{\lii \perp}$, respectively.

To determine the critical exponents from our numerical
data, we consider the scaling relation
\begin{align}
\label{eq:appscale}
A(\alpha,\beta) = \beta^{-\kappa / \nu} f_A \boldsymbol{(} \beta^{1/\nu} (\alpha - \alpha^\mathrm{c}), \beta^{-\omega} \boldsymbol{)} \, ,
\end{align}
where $f_A$ is a universal function, $\alpha$ is a parameter that is tuned across the critical coupling $\alpha^\mathrm{c}$,
$\beta=1/T$ is inverse temperature, $\beta^{-\omega}$ describes subleading corrections to scaling, and $\kappa$ is an exponent that depends on the observable $A$.
For example, the local moment at zero temperature fulfils $m^2_i(\alpha) \propto (\alpha_i - \alpha_i^\mathrm{c})^{2\beta^\prime}$ such that $\kappa = 2\beta^\prime$.
For most of our analysis, we consider RG-invariant observables with $\kappa = 0$
like the normalized correlation length $\xi_i /\beta$
or the correlation ratio $R_i$ defined in Eqs.~\eqref{eq:clength} and \eqref{eq:cratio}, respectively.

For $\kappa=0$ and in the absence of any correction terms, all data sets $A(\alpha,\beta)$ exhibit a common crossing at the critical coupling $\alpha^\mathrm{c}$,
independent of the chosen $\beta$. In this case, it is sufficient to tune $1/\nu$ until all our numerical data collapse onto the universal function $f_A$.
In the presence of subleading corrections,
we extract the pseudocritical couplings
$\alpha^A_\mathrm{pc}(\beta)$ from the crossings between data sets $(\beta, r \beta)$, $r>1$. For $\beta \to \infty$, the pseudocritical couplings converge to the critical coupling
$\propto \beta^{-(1/\nu+\omega)}$. Such a crossing analysis was used to
obtain the fixed-point couplings in Fig.~\ref{fig:FPannihilation} and the phase diagrams shown below, as described in detail in Ref.~\cite{PhysRevLett.130.186701}.
While there are many ways to extract the critical exponent $1/\nu$ from the scaling ansatz in Eq.~\eqref{eq:appscale},
the crossing analysis allows us to define a sliding critical exponent
\begin{align}
\label{eq:appnu}
\frac{1}{\nu^A_\mathrm{pc}(\beta)}
	=
	\frac{1}{\ln r} \ln \left( \frac{dA(\alpha,r \beta)/d\alpha}{dA(\alpha,\beta)/d\alpha} \right)_{\alpha=\alpha^A_\mathrm{pc}(\beta)}
\end{align}
that converges to the true exponent with $\mathcal{O}(\beta^{-\omega})$ corrections and has been used in the study of deconfined criticality \cite{doi:10.1126/science.aad5007}.
In practice, we evaluate the derivatives by fitting each data set with a cubic function near the crossing and use a bootstrapping analysis to estimate the statistical error. Our estimator for the sliding exponent will become useful in Sec.~\ref{Sec:Pseudocriticality}, when we discuss how the slow RG flow within the critical manifold close to the fixed-point collision affects the finite-size scaling at the quantum phase transition.

\section{Phase diagrams and crossover behavior at finite anisotropies\label{Sec:PhaseDiagAni}}

In this section, we use our QMC method to determine the phase diagram of the U(1)-symmetric anisotropic spin-boson model
along different cuts in parameter space at fixed $\alpha_z / \alpha_{xy}$. Moreover, we study the crossover behavior
from the SU(2)-symmetric case towards the fixed points in the U(1)-symmetric plane.

\subsection{Finite-size analysis of the spin susceptibility\label{Sec:FSSsus}}

First, we want to discuss how we determine the fixed-point couplings and phase boundaries presented throughout this work.
It is convenient to consider the
normalized correlation length $\xi_{xy}/\beta$ defined in Eq.~\eqref{eq:clength} which
diverges in the localized phases L1/2, but remains finite in the critical phases CR1/2.
Exactly at the RG fixed points, $\xi_{xy}/\beta$ becomes scale invariant and exhibits
crossings for different temperatures if plotted as a function of the dissipation strength.
While $\xi_{xy}/\beta$ is expected to remain finite throughout the entire critical phase,
it exhibits subleading corrections away from the fixed points. In the same way,
we can use the prediction of the perturbative RG that $\chi_{xy}(T) \propto T^{-s}$ at the critical fixed point (and therefore throughout the critical phase with additional subleading corrections).
Then, $T^s \chi_{xy}$ exhibits the same crossings as $\xi_{xy}/\beta$, but smaller statistical fluctuations
of the QMC estimator make the analysis of $T^s \chi_{xy}$ more precise \cite{PhysRevLett.130.186701}.

\begin{figure}
\includegraphics[width=\linewidth]{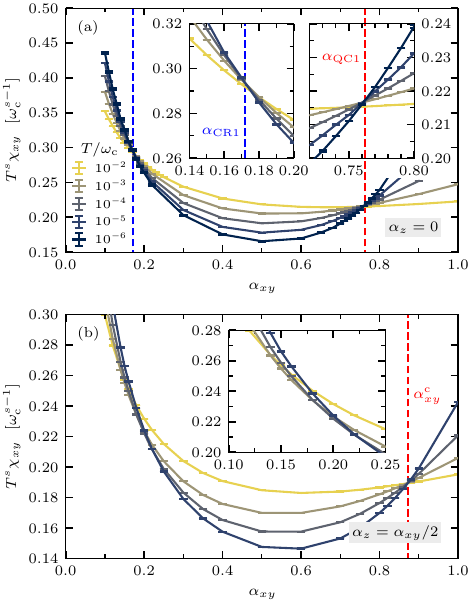}
\caption{%
Finite-temperature analysis of the rescaled susceptibility $T^s \chi_{xy}$ as a function of the dissipation strength $\alpha_{xy}$ for fixed anisotropies (a) $\alpha_z=0$
and (b) $\alpha_z = \alpha_{xy}/2$ at $s=0.8$.
(a) At $\alpha_z=0$,
$T^s \chi_{xy}$ exhibits two clean crossings, as highlighted in the two insets, which correspond to the critical fixed point $\cri$ and the quantum critical fixed point $\qci$ and are marked by the vertical dashed lines.
(b) At $\alpha_z = \alpha_{xy}/2$, we observe a clear crossing at strong couplings, which 
corresponds to the critical coupling $\alpha^\mathrm{c}_{xy}$ between the critical and the localized phase. However, the crossings at weak couplings do not converge to a fixed value, but significantly drift with decreasing temperature (see inset); this is consistent with the absence of a weak-coupling fixed point at finite anisotropies.
}
\label{fig:crossings}
\end{figure}
Figure \ref{fig:crossings}(a) shows $T^s \chi_{xy}$ as a function of $\alpha_{xy}$ for the U(1)-symmetric model at $\alpha_z=0$ and $s=0.8$. We observe two clean crossings which correspond to the weak- and strong-coupling fixed points $\cri$ and $\qci$, respectively. To estimate the precise fixed-point values, we determine the crossings between data pairs $(T, T/10)$ and extrapolate them towards $T\to 0$ using a power-law fitting function; the details of this analysis have been described in Ref.~\cite{PhysRevLett.130.186701} and its Supplemental Material. The extrapolated fixed-point couplings are shown in Fig.~\ref{fig:FPannihilation}(a).

Figure \ref{fig:crossings}(b) shows the same finite-temperature analysis, but for $\alpha_z = \alpha_{xy}/2$.
Again, we find a clean crossing at strong couplings, which, compared to the case of $\alpha_z = 0$ depicted in Fig.~\ref{fig:crossings}(a), has shifted towards larger values of $\alpha_{xy}$. This critical coupling $\alpha_{xy}^\mathrm{c}$ marks the quantum phase transition between the $\cri$ and $\li$ phases to which our system flows under the RG starting from either side of the separatrix, as illustrated in Fig.~\ref{fig:RGflow}(b). However, the clean crossing at weak couplings has dissolved into pairwise intersections that drift substantially as we lower the temperature. This is a direct consequence of the fact that there is no weak-coupling fixed point in the $\alpha_z = \alpha_{xy}/2$ plane, but the system flows towards $\cri$ at $\alpha_z=0$. All in all, our finite-temperature analysis of $T^s \chi_{xy}$ is in excellent agreement with the RG picture discussed in Sec.~\ref{Sec:RG_summary}.

\subsection{Phase diagrams at different anisotropies\label{Sec:phasediagrams}}

Based on our RG analysis in Sec.~\ref{Sec:RG_summary} and our numerical analysis in Sec.~\ref{Sec:FSSsus}, we can determine the phase diagrams of the anisotropic
spin-boson model along different cuts in parameter space.

\begin{figure}
\includegraphics[width=\linewidth]{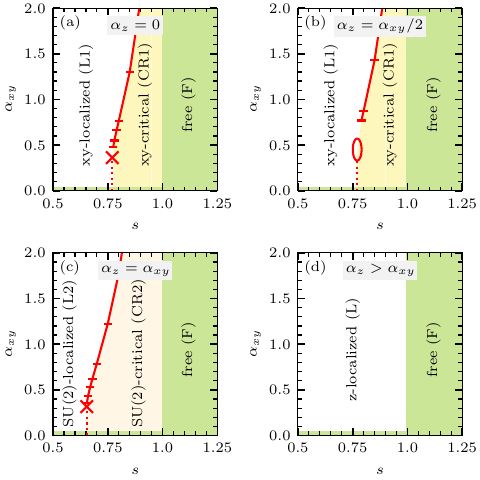}
\caption{%
Phase diagrams as a function of the bath exponent $s$ and the spin-boson coupling $\alpha_{xy}$
for different spin anisotropies (a) $\alpha_z=0$, (b) $\alpha_z=\alpha_{xy}/2$, (c) $\alpha_z = \alpha_{xy}$, and (d) $\alpha_z > \alpha_{xy}$.
The red crosses in (a) and (c) indicate the positions of the fixed-point collisions estimated in Fig.~\ref{fig:FPduality}, whereas the red ellipse in (b) marks the region where the critical coupling 
is supposed to disappear.
}
\label{fig:phasediagrams}
\end{figure}
Figure \ref{fig:phasediagrams}(a) shows the phase diagram of the U(1)-symmetric spin-boson model at $\alpha_z = 0$ as a function of the bath exponent $s$ and the dissipation strength $\alpha_{xy}$. Our system is in the free-spin phase $\f$ for $s>1$ and enters the critical phase $\cri$ for $s<1$, which is destroyed at strong couplings $\alpha > \aqci$ as well as for $s<\si$, \ie, beyond the fixed-point collision. Our QMC results are in good agreement with the phase boundaries obtained from previous MPS studies \cite{PhysRevLett.108.160401, PhysRevB.90.245130, Bruognolo_thesis}.

At $\alpha_z = \alpha_{xy}/2$, the stable phases shown in Fig.~\ref{fig:phasediagrams}(b) are the same as in Fig.~\ref{fig:phasediagrams}(a), only the phase boundary has slightly shifted towards larger couplings. Because the phase structure at $\alpha_z < \alpha_{xy}$ is determined by the fixed points lying in the $\alpha_z = 0$ plane, the critical phase can only exist for $\si < s < 1$. The estimation of $\alpha^\mathrm{c}_{xy}(s = \si)$ is complicated by the slow RG flow near $\si$ and the absence of an appropriate fitting function. However, already for $s \lesssim \si$ we do not find well-defined crossings in $T^s \chi_{xy}$ or $\xi_{xy}/\beta$ anymore that would signal a quantum phase transition.

At $\alpha_z = \alpha_{xy}$, the phase diagram in Fig.~\ref{fig:phasediagrams}(c)
is now determined by the localized and critical fixed points $\lii$ and $\crii$ of the SU(2)-symmetric manifold, in which the critical phase remains stable down to bath exponents of $\sii < \si$ and up to larger couplings $\aqcii > \aqci$. This has important consequences for the properties of the quantum phase transitions tuned through the SU(2)-symmetric plane, as discussed in Sec.~\ref{Sec:QC_anisotropy}.

Eventually, for $\alpha_z > \alpha_{xy}$ shown in Fig.~\ref{fig:phasediagrams}(d), the subohmic regime is governed by the $\lo$ fixed point along the spin $z$ direction.

\subsection{Anisotropy effects on the phase boundary}

\begin{figure}
\includegraphics[width=\linewidth]{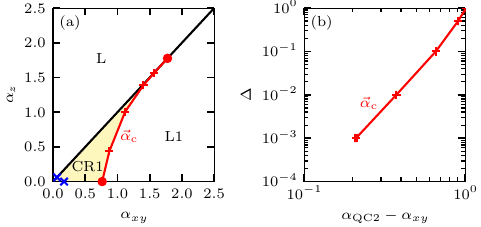}
\caption{%
(a) Phase diagram at $s=0.8$ as a function of the spin-boson couplings $\alpha_{xy}$ and $\alpha_z$.
The blue markers (red circles) indicate the positions of the critical (quantum critical) fixed points.
(b) Evolution of the critical coupling $\vec{\alpha}_\mathrm{c}$ as a function of the distance to $\qcii$ and of the spin anisotropy $\Delta = 1 - \alpha_z / \alpha_{xy}$ on a log-log scale.
}
\label{fig:pds08}
\end{figure}

Figures~\ref{fig:crossings} and \ref{fig:phasediagrams} have already indicated that the critical coupling $\alpha^\mathrm{c}_{xy}$ between the $\cri$ and $\li$ phases at $\alpha_z < \alpha_{xy}$ increases with the spin anisotropy. In Fig.~\ref{fig:pds08}(a), we take a closer look at the evolution of $\vec{\alpha}_\mathrm{c}=(\alpha^\mathrm{c}_{xy}, \alpha^\mathrm{c}_{xy}, \alpha^\mathrm{c}_{z})$ as a function of the anisotropy at fixed $s=0.8$.
It turns out that the critical phase becomes extremely narrow close to $\vec{\alpha}_\qcii$.
To quantify how $\vec{\alpha}_\mathrm{c}$ approaches $\vec{\alpha}_\qcii$, we plot the anisotropy
parameter $\Delta = 1 - \alpha_z / \alpha_{xy}$ at the critical coupling as a function of the
distance to $\aqcii$ in Fig.~\ref{fig:phasediagrams}(b). Our data are consistent
with a power-law convergence towards $\aqcii$, such that its slope
becomes zero when approaching the SU(2)-symmetric manifold.

\subsection{Crossover behavior at finite anisotropies}

For $\alpha_z < \alpha_{xy}$, all phases and their properties are eventually determined
by the fixed points at $\alpha_z = 0$. Here, we study the crossover behavior
from the SU(2)-symmetric manifold towards the U(1)-symmetric fixed points.

\subsubsection{Quantum criticality determined by QC1}

The critical properties at the phase boundary $\vec{\alpha}_\mathrm{c}$ shown in Fig.~\ref{fig:pds08}
are determined by the quantum critical fixed point $\qci$, \ie, the same critical exponents
need to apply to different anisotropies $\Delta$. We consider the normalized correlation length $\xi_{xy}/\beta$ which fulfils the scaling form [\cf Eq.~\eqref{eq:appscale}]
\begin{align}
\label{eq:corrlen}
\xi_{xy} / \beta =
	f^{\qci\parallel} _{\xi_{xy}}\boldsymbol{(}\beta^{1/\nu_{\qci \parallel }} (\alpha_{xy} - \alpha^\mathrm{c}_{xy})\boldsymbol{)} \, .
\end{align}
We tune $\alpha_{xy}$ across the critical coupling $ \alpha^\mathrm{c}_{xy}$ and rescale temperature using the inverse correlation-length exponent $1/\nu_{\qci \parallel}$ at fixed point $\qci$ which lies within the $\alpha_z = 0$ manifold.
At $s=0.8$,
$1/\nu_{\qci \parallel} = 0.1073(15)$ has been determined by QMC \cite{PhysRevLett.130.186701} and is in good agreement with previous MPS results \cite{PhysRevB.90.245130, Bruognolo_thesis}.

\begin{figure}
\includegraphics[width=\linewidth]{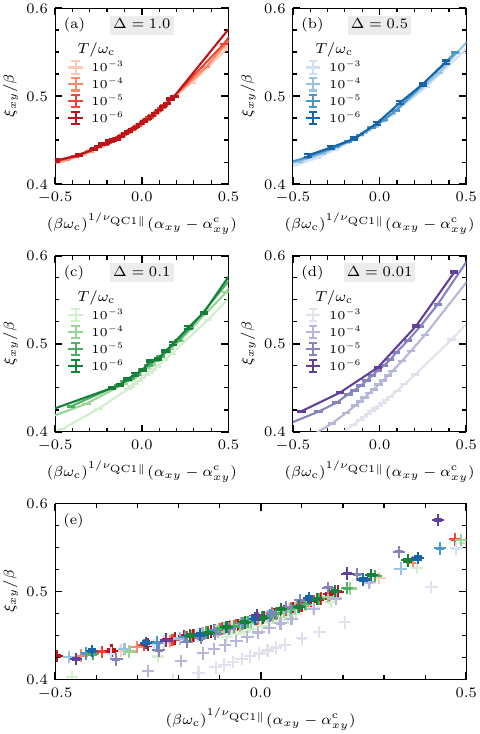}
\caption{%
(a)--(d) Scaling collapse of the correlation length $\xi_{xy}$ for different spin anisotropies
$\Delta
\in \{1.0, 0.5, 0.1, 0.01\}$
and different critical couplings $\alpha^\mathrm{c}_{xy}$,
but with the same correlation-length exponent $\nu_{\qci \parallel} = 0.1073(15)$ of the quantum-critical fixed point $\qci$ \cite{PhysRevLett.130.186701}.
(e) A collection of all data from (a)--(d). Here, $s=0.8$.
}
\label{fig:datacollapseXY08}
\end{figure}
Figure \ref{fig:datacollapseXY08}(a) shows a data collapse of $\xi_{xy}/\beta$
at $\alpha_z = 0$, for which we obtain excellent overlap of all temperature sets that have been considered.
At an intermediate anisotropy of $\Delta = 0.5$ shown in Fig.~\ref{fig:datacollapseXY08}(b),
the agreement is still very good,
only at the highest temperature $T/\omega_\mathrm{c} = 10^{-3}$ our data
starts to diverge slightly earlier from the universal curve.
High-temperature deviations become larger
at $\Delta = 0.1$ shown in Fig.~\ref{fig:datacollapseXY08}(c), but the lowest temperatures still converge
to a universal function. Only at substantially smaller anisotropies of $\Delta=0.01$ depicted in Fig.~\ref{fig:datacollapseXY08}(d), convergence towards the expected critical behavior has not yet been achieved. 

If we collect the data for all anisotropies in one plot, as it is the case in Fig.~\ref{fig:datacollapseXY08}(e), 
we observe that the three data sets for $\Delta \gtrsim 0.1$ seem to converge to a universal function, whereas $\Delta = 0.01$ still shows a substantial drift. Note that it is not required that all curves converge to the same universal function, because we perform our data collapse in the bare couplings $\alpha_{xy}$ at finite anisotropies and not in the scaling variable at the fixed point. The fact that we do not find considerable deviations for $\Delta \gtrsim 0.1$ indicates that such corrections are small in this regime. Nonetheless, at criticality all data sets need to converge to the same value of $\xi_{xy}/\beta$. 

In principle, our crossover analysis can be repeated for other critical exponents and for other bath exponents $s$. Because for $\alpha_z < \alpha_{xy}$  the fixed points of the two-bath spin-boson model determine the critical behavior, we refer to Ref.~\cite{PhysRevB.90.245130} for an extensive discussion of the quantum criticality at these fixed points.

\subsubsection{Crossover of the spin susceptibilities}

\begin{figure}
\includegraphics[width=\linewidth]{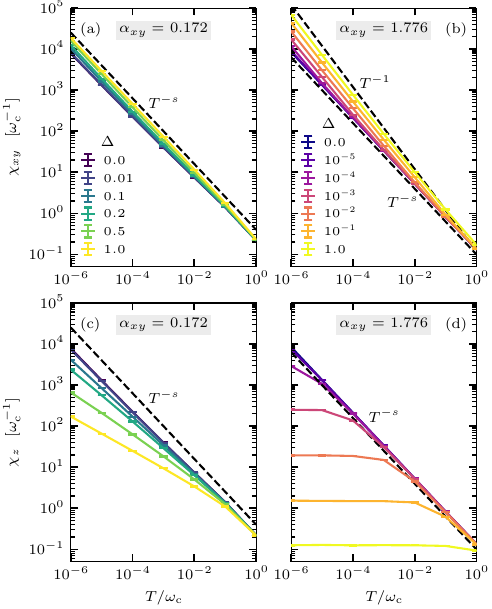}
\caption{%
Temperature dependence of the spin susceptibilities $\chi_{xy}$ and $\chi_z$
for different anisotropies $\Delta$. Our results illustrate the crossover from the SU(2)-symmetric three-bath model at $\Delta=0.0$ to the U(1)-symmetric two-bath model at $\Delta=1.0$ for fixed dissipation strengths (a), (c) $\alpha_{xy} = 0.172 \simeq \acri$ and (b), (d) $\alpha_{xy} = 1.776 \simeq \aqcii$. Black dashed lines indicate the asymptotic behavior $T^{-s}$ and $T^{-1}$, as labeled in the different panels. Here, $s=0.8$.
}
\label{fig:crossover08}
\end{figure}

At small anisotropies $\Delta = 1-\alpha_z / \alpha_{xy}$ and sufficiently high temperatures, signatures of the SU(2)-symmetric fixed points should remain accessible before a crossover towards the U(1)-symmetric physics occurs. Figure~\ref{fig:crossover08}
illustrates this crossover for the spin susceptibilities $\chi_{xy}$ and $\chi_z$ as a function of temperature and for different $\Delta$.

First, we consider a coupling $\alpha_{xy}=0.172$ that is deep within the critical phase and vary $\Delta$; then, the in-plane susceptibility follows a power law $\chi_{xy} \propto T^{-s}$ for all $\Delta \leq 1$, as shown in Fig.~\ref{fig:crossover08}(a). Here, we have chosen  $\alpha_{xy} = \acri$, such that for $\Delta=1$ our path terminates exactly at $\cri$. Therefore, subleading corrections vanish at $\Delta=1$ and lead to a clean power-law behavior, whereas for any $0\leq \Delta <1$ correction terms are still present and $\chi_{xy}$ approaches the $T^{-s}$ behavior only slowly. The out-of-plane susceptibility $\chi_{z}$ shown in Fig.~\ref{fig:crossover08}(c) is expected to follow $T^{-s}$ only at $\Delta=0$ (with subleading corrections) and to perform a crossover towards $\chi_z \propto T^{-\tilde{\eta}_z}$ for any $\Delta > 0$ which is determined by the exponent $\tilde{\eta}_z = 1-\eta_z$ at the fixed point $\cri$ [note that $\eta_z$ has been defined in Sec.~\ref{Sec:pertRG} via the imaginary-time decay of $\chi_z(\tau) \propto \tau^{-\eta_z}$]. At $\Delta=1$, we estimate $\tilde{\eta}_z \approx 0.42$, which is in agreement with previous MPS results on the two-bath spin-boson model \cite{PhysRevB.90.245130}, where the dependence on the bath exponent $s$ has been calculated in more detail and compared with the perturbative result \cite{PhysRevB.66.024427}. For both susceptibilities, we observe that $\Delta=0.01$ is indistinguishable from the SU(2)-symmetric case for $T/\omega_\mathrm{c} \gtrsim 10^{-6}$, whereas for larger anisotropies the crossover temperature increases steadily.
We find that the crossover of the out-of-plane susceptibility $\chi_z$ occurs very slowly and even for $\Delta =0.5$ has not approached $T^{-\tilde{\eta}_z}$. Similar behavior has been observed at $\alpha_z=0$ when tuning $\alpha_{xy}$ away from the stable fixed-point coupling \cite{PhysRevB.105.165129}, so that one could easily misinterpret this behavior as a varying exponent.
Such a slow RG flow is characteristic near the critical fixed points and related to the small inverse correlation-length exponents along both directions, as discussed in Sec.~\ref{Sec:ContinuousCriticality}.

We also study the crossover behavior starting from the quantum critical fixed point $\qcii$ at $\alpha_{xy} = 1.776$, as shown in Figs.~\ref{fig:crossover08}(b) and \ref{fig:crossover08}(d). At $\Delta=0$, $\chi_{xy}$ and $\chi_z$ follow a clean power law $T^{-s}$, whereas for $\Delta > 0$ our system flows towards the $xy$-localized fixed point $\li$ where $\chi_{xy} \propto T^{-1}$ and $\chi_z \to \mathrm{const}$. In contrast to the critical regime analyzed before, where our data for $\Delta=0.01$ were indistinguishable from the SU(2)-symmetric case down to temperatures $T/\omega_\mathrm{c} \approx 10^{-6}$, the same anisotropy already leads to deviations on a scale of $T/\omega_\mathrm{c} \approx 10^{-2}$. At $\alpha_{xy} = 1.776$, it appears that the anisotropy sets the energy scale at which our data starts to diverge from the $\Delta=0$ case. Eventually, we need anisotropies of $\Delta = 10^{-5}$ for our data to become indistinguishable from the isotropic case for all temperatures $T/\omega_\mathrm{c} > 10^{-6}$ considered in Fig.~\ref{fig:crossover08}.
As a result, we would need extremely small anisotropies if we wanted to access the critical exponent $1/\nu_{\qcii \parallel}$ at the $\crii$--$\lii$ transition from finite-temperature measurements, before the system will start to flow towards $\qci$. This is in agreement with our previous discussion of Fig.~\ref{fig:datacollapseXY08}.

All in all, the crossover scales are very different for the two fixed points $\crii$ and $\qcii$. While the properties of the former can be accessed at reasonable anisotropies of $\Delta=0.01$, the latter requires $\Delta=10^{-5}$ for the same precision at comparable temperature scales. 
This suggests that the approximate fixed-point duality is only valid within the high-symmetry manifold, but does not apply perpendicular to it.
Of course, the different scales can change significantly as the bath exponent $s$ approaches the fixed-point collision. Below, we will study the flow away from $\crii$ in more detail.

\section{Phase transitions across the SU(2)-symmetric manifold\label{Sec:QC_anisotropy}}

In this section, we will study how the nontrivial fixed-point structure within the SU(2)-symmetric parameter manifold determines the rich critical behavior driven by anisotropy. In particular, we will characterize the first- and second-order transitions between different localized and critical phases and how the fixed-point annihilation provides a generic mechanism for a weak first-order transition that is not fine tuned. Eventually, we will present a detailed discussion of pseudocritical scaling near the fixed-point collision and provide direct numerical evidence for this scenario.

\subsection{Overview of tunable criticality from an analysis of the order parameter}
\label{Sec:OP}

From our analysis of the possible RG flow diagrams in Sec.~\ref{Sec:RG_summary}, we have a clear picture
of what to expect at the anisotropy-driven quantum phase transition across the high-symmetry manifold at
$\alpha_z = \alpha_{xy}$, as we tune the bath exponent $s$. Here, we will give an overview
of the transition by first looking at the order parameter across the SU(2)-symmetric manifold and then by characterizing the local moment within the high-symmetry plane.

\subsubsection{Local moment across the transition}
\label{Sec:LocalMom}

\begin{figure}
\includegraphics[width=\linewidth]{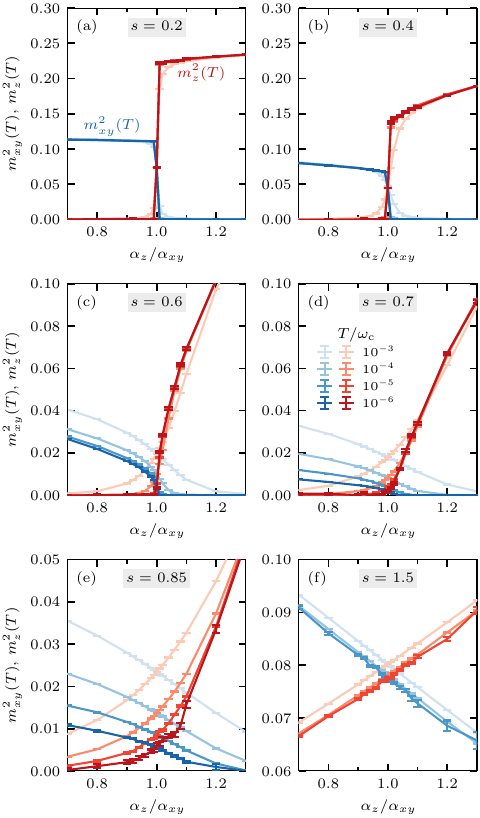}
\caption{%
Finite-temperature scaling of the local moment estimators $m^2_{xy}(T)$ (blue) and $m^2_z(T)$ (red).
Results are shown for various bath exponents $s$ and as a function of the anisotropy $\alpha_z / \alpha_{xy}$.
For all data, we fix $\alpha_{xy} = 0.317 \simeq \alpha^\ast_2$.
}
\label{fig:localmoments}
\end{figure}

Figure~\ref{fig:localmoments} shows a finite-size-scaling analysis
for the local moments $m^2_{xy}(T)$ and $m^2_z(T)$ estimated via Eq.~\eqref{eq:locmom} for different bath exponents $s$.
We fix $\alpha_{xy}=0.317 \simeq \aii$, \ie, we cross the critical manifold at the dissipation strength at which the two intermediate-coupling fixed points collide for $\sii = 0.6540(2)$, and tune the anisotropy parameter $\alpha_z / \alpha_{xy}$ across the transition.

For $s=0.2$ and $0.4$, as shown in Figs.~\ref{fig:localmoments}(a) and \ref{fig:localmoments}(b), we observe a strong first-order transition which occurs between the $\li$ phase at $\alpha_{z}/\alpha_{xy}<1$
and the $\lo$ phase at $\alpha_{z}/\alpha_{xy} > 1$.
The $\li$ phase is characterized by $m^2_{xy} > 0$ and $m^2_z = 0$, whereas in the $\lo$
phase $m^2_{xy} =0$ and $m^2_z > 0$. Exactly at the transition, the two
orders coexist, \ie, $m^2_{xy} = m^2_z > 0$, because the SU(2)-symmetric transition point is governed by the localized fixed point $\lii$.
For $s=0.6$ shown in Fig.~\ref{fig:localmoments}(c), our system still exhibits a first-order transition because we are in the regime $s < \sii$ illustrated in Fig.~\ref{fig:RGflow}(d).
However, the local moment at the transition point is so small that it has not yet converged for the available temperatures. As a result, the system exhibits a weak first-order transition.

For $s=0.7$ shown in Fig.~\ref{fig:localmoments}(d), our RG flow analysis illustrated in Fig.~\ref{fig:RGflow}(c) suggests that we are in the regime $\sii < s < \si$, for which an additional (attractive) critical fixed point $\crii$ exists at $\alpha_z = \alpha_{xy}$. For the parameters chosen in Fig.~\ref{fig:localmoments}(d), we cross the high-symmetry manifold within the basin of attraction of $\crii$. Therefore, $m^2_{xy} = m^2_z = 0$, so that we expect a continuous transition between the $\li$ and $\lo$ phases. We also notice that the order parameter $m^2_{xy} > 0$ is substantially suppressed for $\alpha_z / \alpha_{xy}<1$ because for $s=0.7$ our system is already  close to the fixed-point collision of the U(1)-symmetric model at $\si = 0.7706(1)$, which leads to a slow convergence of the small order parameter.

For $s=0.85$ shown in Fig.~\ref{fig:localmoments}(e), our system falls into the regime $\si < s <1$, for which the schematic RG flow is illustrated in Fig.~\ref{fig:RGflow}(b). There is an additional stable fixed point $\cri$ in the $xy$ plane, such that a stable critical phase can exist for $\alpha_z/\alpha_{xy}<1$. This is exactly the case for
the parameters in Fig.~\ref{fig:localmoments}(e), for which the system exhibits a continuous $\cri$--$\lo$ transition. The convergence of $m^2_{xy}(T) \to 0$ for $\alpha_z / \alpha_{xy} < 1$ is even slower
than for $s=0.7$, where $m^2_{xy}$ still exhibited a small but finite local moment. This can be understood from the temperature dependence of the susceptibility within the critical phase, $\chi_{xy}(T) \propto T^{-s}$;
the convergence of $m^2_{xy}(T) \propto T^{1-s}$ for $T\to 0$ is expected to become slower with increasing $s \to 1$. For the same reasons, the temperature convergence is also slow at $\alpha_z = \alpha_{xy}$.

Finally, Fig.~\ref{fig:localmoments}(f) shows the local moments in the superohmic regime
at $s=1.5$, where the system is always in the free-spin phase $\f$ according to the RG flow
depicted in Fig.~\ref{fig:RGflow}(a). Indeed, we find $m^2_{xy}, m^2_z > 0$ for all finite anisotropies.
In general, their absolute values differ, but they become equal at the high-symmetry point.

\subsubsection{Fixed-point annihilation in the high-symmetry manifold as a mechanism for a weak first-order transition}

In the following, we
take a closer look at the local moment $m^2_\mathrm{SU(2)}$ of the SU(2)-symmetric model,
because it determines the coexistence of the two order parameters $m^2_{xy}$ and $m^2_z$ of the anisotropic system right at the transition. Most importantly, $m^2_\mathrm{SU(2)}$ gives quantitative insight into the extent of the parameter regime in which we can find a weak first-order transition.

\begin{figure}[t]
\includegraphics[width=\linewidth]{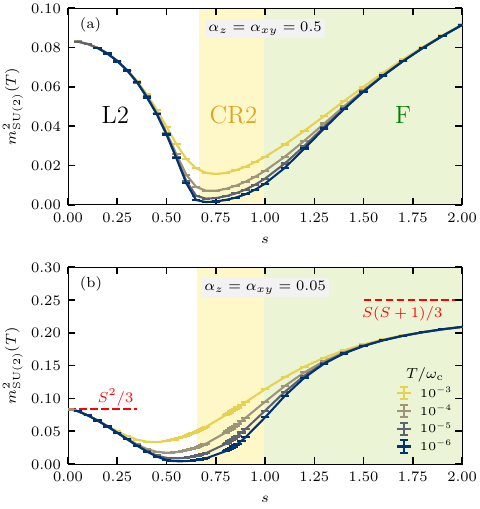}
\caption{%
Temperature convergence of the local moment estimate $m^2_\mathrm{SU(2)}(T)$
as a function of the bath exponent $s$
at dissipation strengths (a) $\alpha_z = \alpha_{xy} = 0.5$
and (b) $\alpha_z = \alpha_{xy} = 0.05$.
The shaded areas correspond to the stable phases $\lii$, $\crii$, and $\f$ of the SU(2)-symmetric model.
The red dashed lines in (b) indicate the local moments of a quantum spin [$m^2_\mathrm{SU(2)} = S(S+1)/3$] and a classical spin [$m^2_\mathrm{SU(2)} = S^2/3$].
Results are taken from the Supplemental Material of Ref.~\cite{PhysRevLett.130.186701}.
}
\label{fig:localmomentSU2}
\end{figure}

Figure \ref{fig:localmomentSU2}(a) shows $m^2_\mathrm{SU(2)}(T)$ at fixed $\alpha_z = \alpha_{xy} = 0.5$,
for which the dissipation strength is chosen to be larger than $\aii = 0.317(1)$, \ie, the coupling at the fixed-point collision.
The local moment remains finite within the $\lii$ and $\f$ phases, but continuously scales to zero when approaching the $\crii$ phase. For the anisotropy-driven transition, we are particularly interested in the vanishing of $m^2_\mathrm{SU(2)}$ when tuning $s$ from the $\lii$ phase towards the $\crii$ phase. In order to get an arbitrarily small order parameter,
we have to tune the system close to the phase boundary. In this case, the weak first-order $\li$--$\lo$ transition is fine tuned.

In Fig.~\ref{fig:localmomentSU2}(b), the bath coupling $\alpha_z = \alpha_{xy} = 0.05$
is chosen to be much smaller than $\aii = 0.317(1)$.
As a result, the local moment is strongly suppressed for $s\lesssim \sii$. A naive extrapolation
suggests that the order parameter is almost zero at $s\approx 0.5$. However, the fixed-point collision
will only occur at $\sii = 0.6540(2)$, leaving a wide parameter range where $m^2_\mathrm{SU(2)}$
is extremely small. Consequently, the weak first-order transition driven through this extended region in parameter space is not fine tuned.
An analysis of the pseudocritical scaling that naturally occurs in this regime will be discussed in Sec.~\ref{Sec:Pseudocriticality}.
A detailed discussion of the SU(2)-symmetric spin-boson model, including further results on the
local moment and the fixed-point annihilation, can be found in Ref.~\cite{PhysRevLett.130.186701}.

Beyond our interest in exotic criticality, the evolution of the local moment as a function of the bath exponent $s$ also gives insight into the quantum-to-classical crossover of a spin coupled to the environment. For $s\to\infty$, the bath density of states is essentially zero and therefore we recover the local moment of a free spin, \ie, $S(S+1)/3$.
By contrast, for $s\to 0$ the dynamical fluctuations of the spin are substantially suppressed by the coupling to the bath, so that the spin gets stuck
in a classical state with a local moment of $S^2/3$; this happens for any dissipation strength \cite{PhysRevLett.130.186701}.

\subsection{Critical properties at fixed points L2 and CR2}

Our preceding study of the order parameter in Sec.~\ref{Sec:OP} can only give a first impression of the phase transitions experienced by the anisotropic spin-boson model.
As follows, we will characterize the critical properties of the first-order and second-order transitions by performing
a detailed finite-size-scaling analysis that gives us access to the critical exponents at both fixed points $\lii$ and $\crii$.

\subsubsection{Finite-size scaling at the first-order transition}
\label{Sec:FirstOrderScaling}

The first-order transition between the two long-range-ordered localized phases $\li$ and $\lo$ is described by the fixed point $\lii$ that is stable within the SU(2)-symmetric manifold but unstable to anisotropy. We have observed in Fig.~\ref{fig:localmoments} that the local-moment order parameter is discontinuous across the transition.
RG predicts that finite-size scaling also holds at this discontinuity fixed point  \cite{PhysRevLett.35.477, PhysRevB.26.2507}.

\begin{figure}
\includegraphics[width=\linewidth]{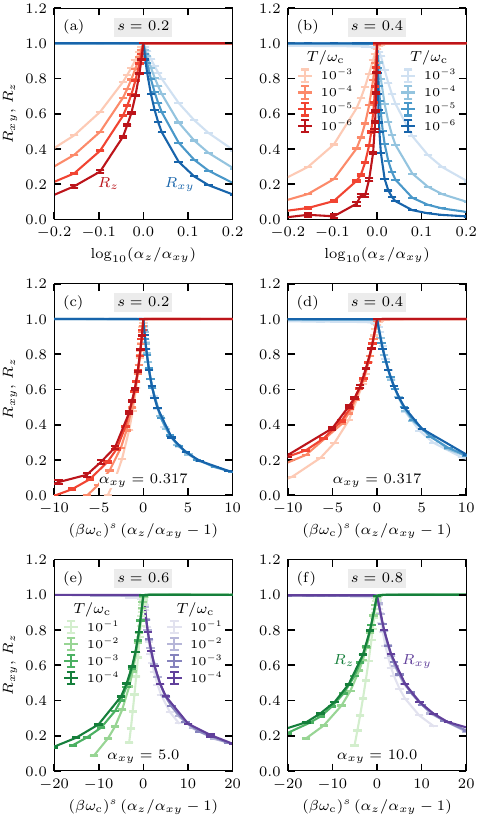}
\caption{%
(a), (b) Finite-temperature analysis of the correlation ratios $R_{xy}$ (blue) and $R_z$ (red)
at $s=0.2$ and $0.4$ as a function of $\alpha_z / \alpha_{xy}$ with $\alpha_{xy}=0.317\simeq \alpha^\ast_2$.
We use a logarithmic scale for the anisotropy, because then the short-range correlations
across the first-order transition become symmetric under
$\alpha_z / \alpha_{xy} \leftrightarrow \alpha_{xy} / \alpha_z$. Note that the data points have not been chosen according to this symmetry.
(c)--(f) Scaling collapse of the correlation ratios for $s\in\{0.2, 0.4, 0.6, 0.8\}$ with $1/\nu_{\lii \perp} = s$. Panels (c) and (d) show the same data as in (a) and (b), whereas for panels (e) and (f) we choose stronger dissipation strengths of $\alpha_{xy}=5.0$ and $10.0$, respectively.
}
\label{fig:cratioanisotropyscaling}
\end{figure}
To test the scaling hypothesis, we consider the correlation ratios $R_{xy}$ and $R_z$ defined in Eq.~\eqref{eq:cratio}, which scale to one (zero) in the corresponding (dis)ordered phase. Figures \ref{fig:cratioanisotropyscaling}(a) and \ref{fig:cratioanisotropyscaling}(b) show $R_{xy}$ and $R_z$ at $s=0.2$ and $0.4$ for the same parameters as in Figs.~\ref{fig:localmoments}(a) and \ref{fig:localmoments}(b), \ie, at a  dissipation strength of $\alpha_{xy} =0.317 \simeq \alpha_2^\ast$.
We plot our data against the logarithm of the anisotropy $\alpha_z / \alpha_{xy}$, because then a symmetry between the short-range correlations of $R_{xy}$ and $R_z$
across $\alpha_z / \alpha_{xy} = 1$ becomes apparent. Together with this symmetry, our data suggest that a finite-size scaling ansatz
\begin{align}
\label{eq:Rscaling}
R_i = f^{\lii\perp}_{R_i} \boldsymbol{(}\beta^{1/\nu_{\lii\perp}} (\alpha_{z}/\alpha_{xy}-1)\boldsymbol{)}
\end{align}
based on Eq.~\eqref{eq:appscale} is valid. For a system with short-range interactions, the theory of discontinuity fixed points predicts that the inverse correlation-length exponent is given by the spatial dimension \cite{PhysRevLett.35.477, PhysRevB.26.2507}. For our spin impurity with long-range retarded interactions, we find that the critical exponent is given by the bath exponent $s$, \ie,
\begin{align}
\label{eq:invnuL2}
1/\nu_{\lii\perp} = s \, .
\end{align}
For our data at $s=0.2$ and $0.4$, the corresponding data collapses are shown in Figs.~\ref{fig:cratioanisotropyscaling}(c) and \ref{fig:cratioanisotropyscaling}(d). We observe excellent scaling behavior using the exponent given in Eq.~\eqref{eq:invnuL2}.
However, such a scaling collapse will fail for our data in Fig.~\ref{fig:localmoments}(c) because at $s=0.6$ the RG flow towards $\lii$ becomes extremely slow near the fixed-point collision. 
Therefore, we probe the first-order scaling hypothesis at $s=0.6$ and $0.8$ in the strong-coupling regime with $\alpha_{xy}=5.0$ and $10.0$, as depicted in Figs.~\ref{fig:cratioanisotropyscaling}(e) and \ref{fig:cratioanisotropyscaling}(f), respectively. In this limit, our system quickly flows to the strong-coupling fixed points and we can probe an excellent scaling collapse again.

We note that the data collapse at the first-order transition also occurs for the
normalized correlation lengths $\xi_{xy}/\beta$ and $\xi_{z}/\beta$. However, for $\xi_{xy}$ ($\xi_z$) it
can only be observed for $\alpha_z > \alpha_{xy}$ ($\alpha_z < \alpha_{xy}$), \ie, in the corresponding
disordered phase where the correlation length is finite. In the $xy$ ($z$) ordered phase, $\xi_{xy}/\beta$
($\xi_z / \beta$) diverges with $\beta$. Plotting the correlation ratio \eqref{eq:cratio} in Fig.~\ref{fig:cratioanisotropyscaling} has the advantage that it converges to one in the ordered phase
and therefore hides this issue.

\subsubsection{Nature of the continuous phase transitions}
\label{Sec:ContinuousCriticality}

The existence of the SU(2)-symmetric critical phase $\crii$ for $s>s^\ast_2$
renders the quantum phase transition between the $xy$ and $z$ phases continuous. 
For $s^\ast_2<s < s^\ast_1$, this transition is between the ordered $\li$ and $\lo$ phases,
whereas for $s^\ast_1<s<1$ it is between the critical $\cri$ phase and the $\lo$ phase.
In the following, we characterize these transitions via their critical exponents.

\begin{figure}[t]
\includegraphics[width=\linewidth]{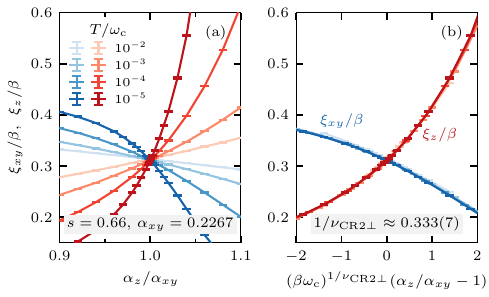}
\caption{%
(a)~Normalized correlation lengths $\xi_{xy}/\beta$ (blue) and $\xi_z/\beta$ (red) as a function of the anisotropy $\alpha_z / \alpha_{xy}$ and for different temperatures. (b) Scaling collapse using the inverse correlation-length exponent $1/\nu_\mathrm{\crii\perp}\approx0.333(7)$. Here, $s=0.66$ and $\alpha_{xy}=0.2267 \simeq \acrii$, so that the transition occurs at $\crii$.
}
\label{fig:datacollapse066}
\end{figure}

Figure~\ref{fig:datacollapse066}(a) shows the normalized correlation lengths $\xi_{xy}/\beta$
and $\xi_z/\beta$ across the phase transition at $s=0.66$ and $\alpha_{xy}=0.2267 \simeq \acrii$. We select $s \gtrsim \sii$ close to the fixed-point annihilation and tune the anisotropy right through the fixed point $\crii$ at $\alpha_z / \alpha_{xy}=1$ to minimize scaling corrections within the critical manifold. For different temperature sets, our data exhibits a common crossing at $\alpha_z = \alpha_{xy}$. Based on Eq.~\eqref{eq:appscale}, we use a scaling ansatz
\begin{align}
\label{eq:xiscaling}
\xi_i / \beta = f^{\crii\perp}_{\xi_i} \boldsymbol{(}\beta^{1/\nu_{\crii\perp}} (\alpha_{z}/\alpha_{xy}-1)\boldsymbol{)} 
\end{align}
and the inverse correlation-length exponent $1/\nu_\mathrm{\crii\perp} = 0.333(7)$ leads to an excellent data collapse of $\xi_{xy}$ and $\xi_z$ across the second-order transition, as illustrated in Fig.~\ref{fig:datacollapse066}(b).
Details on how we extract the correlation-length exponent across $\crii$ from our numerical data can be found in Appendix~\ref{App:getexp}.

\begin{figure}[t]
\includegraphics[width=\linewidth]{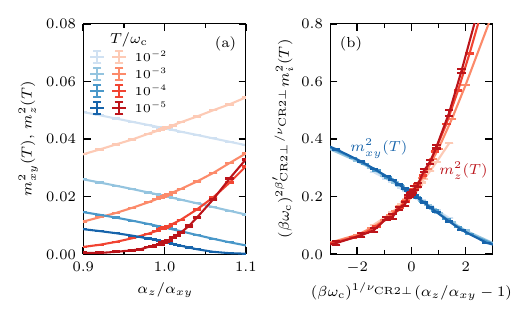}
\caption{%
(a) Finite-temperature estimates of the local moments $m^2_{xy}(T)$ (blue) and $m^2_z(T)$ (red) as a function of the anisotropy $\alpha_z / \alpha_{xy}$ and for different temperatures.
(b) Scaling collapse using the inverse correlation-length exponent $1/\nu_\mathrm{\crii\perp}\approx0.333(7)$ estimated in Fig.~\ref{fig:datacollapse066} and $\beta^\prime_\mathrm{\crii\perp}$ obtained via Eq.~\eqref{eq:hyper}.
Here, $s=0.66$ and $\alpha_{xy}=0.2267\simeq \acrii$, so that we tune the system right through $\crii$ at the transition.
}
\label{fig:datacollapse066_mloc}
\end{figure}

For the same parameters as in Fig.~\ref{fig:datacollapse066}, we also perform a finite-size-scaling analysis of the local-moment estimates $m^2_{xy}(T)$ and $m^2_z(T)$, as shown in Fig.~\ref{fig:datacollapse066_mloc}(a).
The scaling form [\cf Eq.~\eqref{eq:appscale}]
\begin{align}
m^2_i(T) = & \beta^{-2 \beta^\prime_\mathrm{\crii\perp} / \nu_\mathrm{\crii\perp}}
f^{\crii\perp}_{m^2_i}\boldsymbol{(}\beta^{1/\nu_{\crii\perp}} (\alpha_{z}/\alpha_{xy}-1)\boldsymbol{)}
\label{eq:scaling_mloc}
\end{align}
allows us to extract the magnetization exponent $\beta^\prime_\mathrm{\crii\perp}$ from the temperature dependence of $m^2_i(T)$,
given that we have already determined $1/\nu_{\crii\perp}$ from Fig.~\ref{fig:datacollapse066}.
For $\alpha_z=\alpha_{xy}$, the scaling form
becomes
$
m^2_i(T;\vec{\alpha}=\vec{\alpha}_\crii) \propto T^{2\beta^\prime_\mathrm{\crii\perp} / \nu_\mathrm{\crii\perp}}
$.
At the critical fixed point $\crii$, the exact low-temperature behavior of the spin susceptibility $\chi_i(T) \propto T^{-s}$ fixes the ratio of the two critical exponents
\begin{align}
\label{eq:hyper}
\frac{\beta^\prime_\mathrm{\crii\perp}}{\nu_\mathrm{\crii\perp}} = \frac{1-s}{2} \, .
\end{align}
Note that the same hyperscaling
relation holds for the in-plane exponents at $\qci$ and $\qcii$ \cite{PhysRevB.90.245130, PhysRevLett.130.186701}.
If we choose the magnetization exponent according to Eq.~\eqref{eq:hyper}, we obtain a good data collapse for $m^2_i(T)$, as demonstrated in Fig.~\ref{fig:datacollapse066_mloc}(b). 

\begin{figure}[t]
\includegraphics[width=\linewidth]{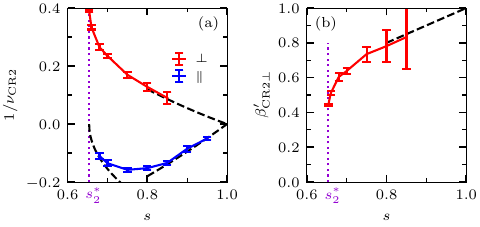}
\caption{%
(a) Inverse correlation-length exponents at the fixed point $\crii$ within ($\parallel$) and perpendicular to ($\perp$) the SU(2)-symmetric manifold as a function of the bath exponent $s$.
The black dashed lines indicate the predictions \eqref{eq:exp_nupara} and \eqref{eq:exp_nu} of the perturbative RG  \cite{PhysRevB.66.024427} as well as the asymptotic behavior \eqref{eq:nupara_FPA} for $s\to\sii$ \cite{PhysRevLett.130.186701}.
(b) Magnetization exponent according to Eq.~\eqref{eq:hyper} with the prediction \eqref{eq:exp_beta} for $s\to 1$.
}
\label{fig:exponents}
\end{figure}

Figure~\ref{fig:exponents} shows $1/\nu_\mathrm{\crii \perp}$ and $\beta^\prime_\mathrm{\crii\perp}$ as a function of the bath exponent $s$.
Both critical exponents are finite at the coordinates $(s^\ast_2,\alpha^\ast_2)$ of the fixed-point collision.
With increasing $s$, $1/\nu_\mathrm{\crii \perp}$ steadily decreases, whereas $\beta^\prime_\mathrm{\crii\perp}$ increases. 
The evolution of the two exponents appears continuous in $s$ and does not take notice of the change in the $xy$ phase
from $\li$ to $\cri$ at $\si=0.7706(1)$; because the critical properties at the anisotropy-driven transition are only defined by the local properties of fixed point $\crii$, this is also not expected.
For $s\to 1$, both exponents approach the prediction of the weak-coupling
perturbative RG \cite{PhysRevB.66.024427} (\cf Appendix~\ref{App:pRG}), \ie,
\begin{gather}
1/\nu_\mathrm{\crii\perp} = \frac{1-s}{2} + \frac{(1-s)^2}{2}
+ \mathcal{O}[(1-s)^3] \, ,
\label{eq:exp_nu}
\\
\beta^\prime_\mathrm{\crii\perp} = 1- (1-s) + \mathcal{O}[(1-s)^2] \, .
\label{eq:exp_beta}
\end{gather}
Note that $\beta^\prime_\mathrm{\crii\perp}$ is obtained from  Eq.~\eqref{eq:exp_nu} using Eq.~\eqref{eq:hyper}.
In particular $1/\nu_\mathrm{\crii \perp} \to 0$ and $\beta^\prime_\mathrm{\crii\perp} \to 1$ for $s\to 1$, consistent with
our numerical data in Fig.~\ref{fig:exponents}.

To complete our analysis of $\crii$, Fig.~\ref{fig:exponents}(a) also contains the inverse correlation-length exponent $1/\nu_{\crii \parallel}$ within the SU(2)-symmetric critical manifold. It can be extracted in the same way as the out-of-plane exponent, we only need to perform our scaling analysis at $\alpha_z = \alpha_{xy}$.
For $s\to 1$, the in-plane exponent approaches the prediction of the perturbative RG \cite{PhysRevB.66.024427},
\begin{align}
1/\nu_\mathrm{\crii\parallel} = -(1-s) + \frac{(1-s)^2}{2}
+ \mathcal{O}[(1-s)^3] \, .
\label{eq:exp_nupara}
\end{align}
For $s \to \sii$, the analyticity of the beta function close to the fixed-point annihilation requires
[\cf Eq.~\eqref{eq:RGquadratic}]
\begin{align}
\label{eq:nupara_FPA}
1/\nu_\mathrm{\crii\parallel} = - C_{\crii \parallel} \sqrt{s-\sii}
+ \mathcal{O}(s-\sii)
 \, .
\end{align}
The proportionality constant $C_{\crii \parallel}  = 0.72(2)$ has been obtained in Ref.~\cite{PhysRevLett.130.186701} by fitting QMC results for $1/\nu_\mathrm{\qcii\parallel}$ to the form of Eq.~\eqref{eq:nupara_FPA}. Because $1/\nu_\mathrm{\qcii\parallel}$ and $1/\nu_\mathrm{\crii\parallel}$ 
must have the same leading behavior near the fixed-point collision, we can directly transfer this result.
Figure~\ref{fig:exponents}(a) confirms that Eq.~\eqref{eq:nupara_FPA} is consistent with our numerical data.
As expected from the approximate fixed-point duality, the absolute numerical values for  $1/\nu_\mathrm{\crii\parallel}$ are close to the ones for  $1/\nu_\mathrm{\qcii\parallel}$ obtained in Ref.~\cite{PhysRevLett.130.186701}.

Finally, we want to note that $1/\nu_\mathrm{\crii\perp}$ seems to exhibit the same nonanalytic behavior for $s\to\sii$ as $1/\nu_\mathrm{\crii\parallel}$.
If we shift Eq.~\eqref{eq:nupara_FPA} by $1/\nu_\mathrm{\crii\perp}(s=\sii) = 0.390(5)$, the analytic prediction fits
the numerical data for $1/\nu_\mathrm{\crii\perp}$ perfectly [we do not show this fit in Fig.~\ref{fig:exponents}(a) because the curves fully overlap].
This equivalence also occurs in the perturbative RG result discussed in Appendix~\ref{App:pRG}, although it goes beyond its range of validity.

\subsection{Pseudocriticality on both sides of the fixed-point collision}
\label{Sec:Pseudocriticality}

Until now, we have studied the critical properties of the spin-boson model only in setups
in which the anisotropy was tuned right through the $\crii$ fixed point or deep in the localized phase
where the system quickly flows to $\lii$. Even for the $\cri$--$\li$ transition at $\alpha_z < \alpha_{xy}$,
the flow towards $\alpha_z = 0$ was rather fast. Usually, it is expected that the component of the RG flow that lies within the critical manifold converges quickly towards its attractive fixed point. As a result, one will probe the same critical exponents at any intersection with the critical manifold.
In the vicinity of the fixed-point collision, this in-plane flow can become extremely slow, such that we do not probe universal behavior but get stuck in a pseudocritical regime. In the following, we want to study this regime in detail for the anisotropic spin-boson model.

\subsubsection{Renormalization-group flow near the fixed-point collision}

\begin{figure*}[t]
\includegraphics[width=\linewidth]{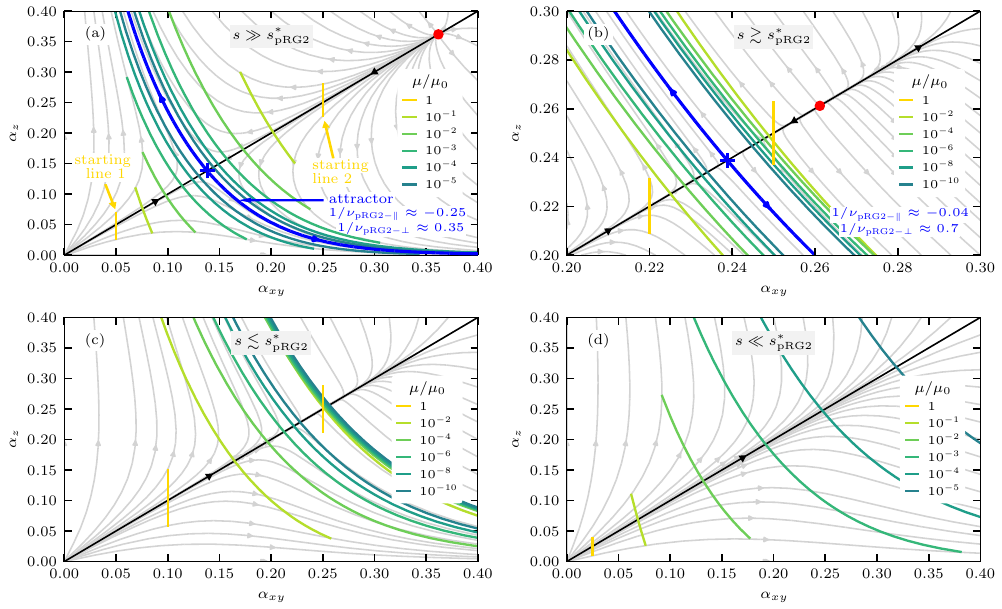}
\caption{%
RG flow diagrams for the two-loop beta function in Eq.~\eqref{eq:beta} as a function of $\alpha_{xy}$ and $\alpha_z$ for different bath exponents $s$
across the fixed-point annihilation, which occurs at $s^\ast_\mathrm{pRG2} = 0.5$ and $\alpha_{xy}=\alpha_z = \alpha^\ast_\mathrm{pRG2} = 0.25$.
Blue crosses (red circles) indicate the stable (unstable) SU(2)-symmetric fixed points which are present for (a) $s=0.6$ and (b) $s=0.501$, but have disappeared for (c) $s=0.499$ and (d) $s=0.4$. We consider different vertical cuts (yellow lines) across $\alpha_z = \alpha_{xy}$ and how they evolve under the RG as the reference scale $\mu / \mu_0$ is reduced (green lines). For $s> s^\ast_\mathrm{pRG2}$, these curves eventually converge to the blue line, which identifies the RG flow from the stable fixed point $\mathrm{pRG}2-$ towards its relevant perturbation. In the vicinity of $\mathrm{pRG}2-$, the RG flow is is characterized by the in- and out-of-plane inverse correlation-length exponents $1/\nu_{\mathrm{pRG}2-\parallel}$ and $1/\nu_{\mathrm{pRG}2-\perp}$, which can lead to an extremely slow RG flow along the diagonal close to the fixed-point collision in (b).
For  $s < s^\ast_\mathrm{pRG2}$, these cuts will eventually flow towards infinity, but for $s \lesssim s^\ast_\mathrm{pRG2}$ shown in (c) there is a regime around $\alpha_{xy} = \alpha_{z} = \alpha^\ast_\mathrm{pRG2}$ with an extremely slow RG flow in $\absolute{s-s^\ast_\mathrm{pRG2}}$.
The consequences of the slow RG flow are discussed in the main text. Further details on the solution of the two-loop beta function \eqref{eq:beta} can be found in Appendix~\ref{App:pRG}.
}
\label{fig:RG_perturbative_st_evolution}
\end{figure*}

Before we continue with the discussion of our numerical results, we first want to review some properties that hold near the fixed-point collision and help us to get a better understanding of the RG flow within and perpendicular to the critical manifold. The approximate beta function close to the collision, as stated in Eq.~\eqref{eq:RGquadratic} and solved in Appendix~\ref{App:RGquad}, can give us a quantitative idea of the RG flow within the critical manifold, but does not contain any information about the perpendicular RG flow. Therefore, we also study the RG flow diagrams of the weak-coupling beta function in Eq.~\eqref{eq:beta}, which contain the two sequential fixed-point annihilations of our model. Although the perturbative beta function does not have  predictive power at strong couplings, the qualitative behavior that is just tied to the existence of a 
fixed-point collision remains reliable. Details on the perturbative RG solution are summarized in Appendix~\ref{App:pRG}. Moreover, our numerical estimates for the in- and out-of-plane correlation-length exponents
in Fig.~\ref{fig:exponents}(a) give us quantitative information about the RG flow near $\crii$.

Figure \ref{fig:RG_perturbative_st_evolution} depicts the RG flow of the beta function \eqref{eq:beta} as a function of $\alpha_{xy}$ and $\alpha_z$ for different bath exponents $s$ on both sides of the fixed-point annihilation. In the context of Fig.~\ref{fig:RG_perturbative_st_evolution}, we call the fixed points within the SU(2)-symmetric manifold $\mathrm{pRG}2\pm$
(as they are obtained from the perturbative RG) and the fixed-point collision occurs at $s^\ast_\mathrm{pRG2}$.
For $s \gg s^\ast_\mathrm{pRG2}$ shown in Fig.~\ref{fig:RG_perturbative_st_evolution}(a), we consider two cuts in parameter space (yellow lines) that cross the critical manifold in some distance to $\mathrm{pRG}2-$.
We assume that our system starts its RG flow on these cuts at the reference scale $\mu_0$ and gets attracted towards the fixed point as we reduce the RG scale $\mu$. During this process, the vertical line deforms, gets stretched out, and eventually converges to the attractor (blue line) that is spanned by the RG flow starting from the fixed point $\mathrm{pRG}2-$ and evolving along the direction of its relevant perturbation. Close to the fixed point, the direction and speed of the RG flow is determined by the in-plane and out-of-plane inverse correlation-length exponents $1/\nu_{\mathrm{pRG2}-\parallel} < 0$ and $1/\nu_{\mathrm{pRG2}-\perp} > 0$, respectively. Along its characteristic directions, the distance to the fixed point $\Delta \bar{\alpha}$ scales as
\begin{align}
\label{eq:RGsc1}
\Delta \bar{\alpha}(\mu)
	\propto
	(\mu/\mu_0)^{-1/\nu}
\end{align}
and therefore decreases (increases) along the $\parallel$ ($\perp$) direction. The speed of the RG flow is determined by the absolute values of $1/\nu$; in Fig.~\ref{fig:RG_perturbative_st_evolution}(a) both exponents are of the same order. However, $1/\nu_{\mathrm{pRG2}-\parallel} \approx -0.25$ is still rather small, so that the flow towards the attractor takes several orders of magnitude in $\mu / \mu_0$, which will become visible in the subleading corrections in our finite-size-scaling analysis. Once our system has converged to this attractor, we are able to measure the perpendicular correlation-length exponent along its direction through the fixed point.

In Fig.~\ref{fig:RG_perturbative_st_evolution}(b), we have tuned this scenario close to the fixed-point collision.
In the vicinity of the fixed point, the scaling form \eqref{eq:RGsc1} is still valid, but the relative scales of the two correlation-length exponents have changed dramatically. The perpendicular RG flow is even faster than before so that the orientation of our vertical line quickly turns parallel to the attractor, but the parallel ($\parallel$) flow towards the fixed point becomes so slow that within several orders of magnitude in $\mu / \mu_0$ we are still far from convergence towards the $\mu \to 0$ limit. While in this regime the RG flow still follows Eq.~\eqref{eq:RGsc1}, it becomes substantially different if we start the RG flow right between the two fixed points where both the beta function and its derivative are close to zero.
Then, the covered distance from $\bar{\alpha}^\ast$ within the critical manifold, \ie, $\Delta\bar{\alpha}_\parallel(\mu) = \absolute{\bar{\alpha}_\parallel(\mu) - \bar{\alpha}^\ast}$, scales as
\begin{align}
\label{eq:RGsc2}
\Delta\bar{\alpha}_\parallel(\mu)
	= a
	\absolute{s-s^\ast} \ln(\mu_0/\mu)  
	\, ,
\end{align}
which is valid in a small interval $\Delta \bar{\alpha}_\parallel(\mu) \ll \sqrt{a \absolute{s-s^\ast}/b}$
in which the beta function can be considered constant
[for a derivation from Eq.~\eqref{eq:RGquadratic} see Appendix~\ref{App:RGquad}]. Because
of this logarithmic scale dependence, the RG flow gets stuck between the two fixed points and only slowly moves forward. In Fig.~\ref{fig:RG_perturbative_st_evolution}(b), our RG flow does not even get close to the attractor. As a result, any numerical simulation starting in this regime will not probe the real critical exponent at the fixed point, but just a local value that drifts extremely slowly toward its expected value. This phenomenon is known as pseudocriticality.
However, the pseudocritical RG flow is limited to a small range in parameter space.

Right after the fixed-point annihilation, the RG flow is illustrated in Fig.~\ref{fig:RG_perturbative_st_evolution}(c)
and only contains the flow from the free fixed point at zero coupling to the corresponding localized fixed points at infinite coupling. However, at the coordinates where the fixed points had collided, the beta function and its derivative are still very small, so that the same pseudocritical flow as in Eq.~\eqref{eq:RGsc2} holds there as well.
In contrast to the previous case, the pseudocritical flow affects all parameters $\bar\alpha_{xy} = \bar\alpha_z \leq \bar\alpha^\ast$ on their flow towards the infinite-coupling fixed point. Then, the total time to flow to the localized fixed point is mainly determined by the pseudocritical region, such that the RG scale $\mu/\mu_0 = \exp(-c/\sqrt{\absolute{s-s^\ast}})$ is exponentially suppressed close to $s^\ast$. Pseudocritical behavior plus the independence of the initial coupling defines the quasiuniversal regime. One of its consequences is the suppression of the order parameter for $s\lesssim s^\ast$ and weak couplings, as observed in Fig.~\ref{fig:localmomentSU2}(b) and Ref.~\cite{PhysRevLett.130.186701}. Although the pseudocritical regime near $\bar{\alpha}^\ast$ is supposed to determine the asymptotic behavior of the RG flow, Fig.~\ref{fig:RG_perturbative_st_evolution}(c) illustrates that close to $s^\ast$ it still takes a considerable amount of RG time to get into this regime.

Eventually, far beyond the fixed-point annihilation, the RG flow speeds up again upon approaching the localized fixed point. This is illustrated in Fig.~\ref{fig:RG_perturbative_st_evolution}(d).

\subsubsection{Numerical results in the pseudocritical regime}

In the following, we provide direct numerical evidence of pseudocritical scaling in the vicinity of the fixed-point collision and how it affects the quantum critical behavior at the anisotropy-driven $\li$--$\lo$ transition.

We start our discussion with the pseudocritical regime at $s>\sii = 0.6540(2)$. To this end, we reconsider our finite-size-scaling analysis at $s=0.66$, but this time we do not tune the anisotropy through $\crii$ at fixed $\alpha_{xy}=0.2267\simeq \acrii$, as it was done in Fig.~\ref{fig:datacollapse066}, but at $\alpha_{xy}=0.317\simeq \aii$ right between the two intermediate-coupling fixed points. 
\begin{figure}[t]
\includegraphics[width=\linewidth]{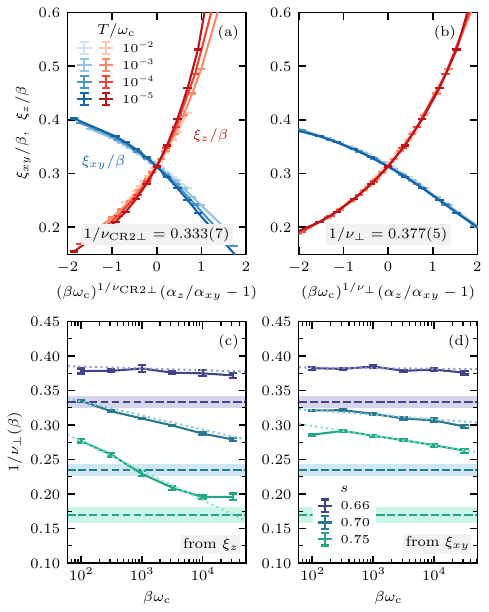}
\caption{%
Pseudocritical scaling before the fixed-point collision ($s>\sii$) at $\alpha_{xy}=0.317\simeq \alpha_2^\ast$.
(a) Insufficient scaling collapse of $\xi_{xy}/\beta$ (blue) and $\xi_z/\beta$ (red) at $s=0.66$ using the inverse correlation-length exponent $1/\nu_{\crii\perp} = 0.333(7)$ at the fixed point $\crii$, as determined in Fig.~\ref{fig:datacollapse066}.
(b) Scaling collapse of the same data, but with $1/\nu_{\crii\perp} = 0.377(5)$.
(c), (d) Drift of the pseudocritical exponent $1/\nu_\perp(\beta)$ with inverse temperature $\beta$,
as obtained from $\xi_{z}$ (left) and $\xi_{xy}$ (right).
Drifting exponents were calculated at the crossing points of data pairs $(\beta, r \beta)$ with $r=\sqrt{10}$ using Eq.~\eqref{eq:appnu}. Dashed lines indicate the exponents calculated right at the fixed point $\crii$, as given in Fig.~\ref{fig:exponents}(a).
Dotted lines represent fits to the functional form $1/\nu_\perp(\beta) = c_1 + c_2 \absolute{s - \sii} \log_{10}(\beta \omega_\mathrm{c})$. To test the linear dependence of the slope on the distance to the fixed-point collision, we have fitted $c_2$ for $s=0.75$ and used the same value of $c_2$ for the fits at smaller $s$.
}
\label{fig:datacollapse066_shifted}
\end{figure}
Figure \ref{fig:datacollapse066_shifted}(a) shows the data collapse that is obtained with the critical exponent $1/\nu_{\crii\perp}\approx0.333(7)$ of $\crii$; apparently, the resulting data collapse is not good.
On the other hand, Fig.~\ref{fig:datacollapse066_shifted}(b) shows an excellent data collapse
for $1/\nu_{\perp}\approx 0.377(5)$. This exponent differs significantly from the one at the fixed point and we do not observe any visible drift of the exponent in the temperature range $\beta \omega_\mathrm{c} \in [10^{2}, 10^{5}]$, as demonstrated in Figs.~\ref{fig:datacollapse066_shifted}(c) and \ref{fig:datacollapse066_shifted}(d). The absence of any visible drift over three orders of magnitude is direct evidence for the extremely slow RG flow in the pseudocritical regime between the two fixed points. Figures \ref{fig:datacollapse066_shifted}(c) and \ref{fig:datacollapse066_shifted}(d) also show the drift of $1/\nu_\perp(\beta)$ for $s=0.7$ and $0.75$.
For these two bath exponents, which are further away from the fixed-point collision, we can observe a finite drift  
of $1/\nu_\perp(\beta)$, but results are still far from convergence to the true critical exponents, as they were obtained right at the fixed point $\crii$ and given in Fig.~\ref{fig:exponents}(a).

\begin{figure}[t]
\includegraphics[width=\linewidth]{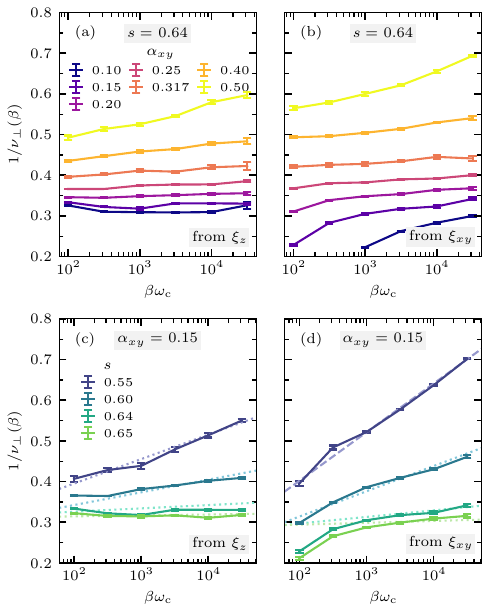}
\caption{%
Drift of the pseudocritical exponent $1/\nu_{\perp}(\beta)$ with inverse temperature $\beta$
after the fixed-point collision ($s < \sii$). We compare exponents obtained from $\xi_{z}$ (left) and $\xi_{xy}$ (right)
(a), (b) at fixed $s=0.64$ for different $\alpha_{xy}$ and (c), (d) at fixed $\alpha_{xy}$ for different $s$.
Drifting exponents were calculated at the crossing points of data pairs $(\beta, r \beta)$ with $r=\sqrt{10}$ using Eq.~\eqref{eq:appnu}. Dotted lines in (c) and (d) represent fits to the functional form $1/\nu_\perp(\beta) = c_1 + c_2 \absolute{s - \sii} \log_{10}(\beta \omega_\mathrm{c})$. To test the linear dependence of the slope on the distance to the fixed-point collision, we have fitted $c_2$ for $s=0.55$ and used the same value of $c_2$ for the fits at larger $s$.
}
\label{fig:nudrift_below}
\end{figure}

Figure \ref{fig:nudrift_below} shows the drift of the pseudocritical exponent $1/\nu_\perp(\beta)$ after the fixed-point annihilation ($s< \sii$). 
First, we fix $s=0.64$ and track the drift of the exponent starting from different couplings $\alpha_{xy}$.
In Fig.~\ref{fig:nudrift_below}(a), where $1/\nu_\perp(\beta)$ is extracted from $\xi_z$, we only observe a very slow drift (if at all) of the exponent for $\alpha_{xy} \lesssim 0.25$. Only for $\alpha_{xy} \gtrsim \aii$, the drift becomes significantly stronger and its slope steadily increases with increasing $\alpha_{xy}$. If we extract $1/\nu_\perp(\beta)$
from $\xi_{xy}$, as shown in Fig.~\ref{fig:nudrift_below}(b), we observe a slow drift for $\alpha_{xy} \lesssim 0.25$, with a slope that is comparable for all data sets, whereas the slope increases again for $\alpha_{xy} \gtrsim \aii$. We note that there is a stronger drift of $1/\nu_\perp(\beta)$ for small $\alpha_{xy}$ and small $\beta$ which can have several reasons: On the one hand, the crossings of data sets $(\beta, r\beta)$ converge slower to the exact critical coupling at $\alpha_z  = \alpha_{xy}$ than for stronger couplings (\cf Appendix~\ref{App:getexp}). On the other hand, our starting values are further away from $\aii$, which could lead to additional correction terms.
Our results in Figs.~\ref{fig:nudrift_below}(a) and \ref{fig:nudrift_below}(b)
are consistent with our expectation that the slope of the drift hardly changes for $\alpha_{xy} \lesssim \aii$ because the RG flow is dominated by the pseudocritical regime at $\alpha_{xy} \approx \aii$, whereas it starts to increase once we have passed this regime.

Our preceding RG analysis predicts that in the pseudocritical regime the in-plane RG flow in Eq.~\eqref{eq:RGsc2} depends on the distance  to the fixed-point collision. To test how this affects the drift of the pseudocritical exponent, we repeat our analysis at fixed $\alpha_{xy}=0.15$ and tune the bath exponent $s < \sii$, as shown in Figs.~\ref{fig:nudrift_below}(c) and \ref{fig:nudrift_below}(d), but also take another look at the drift at fixed $\alpha_{xy}\simeq \aii$ and tune $s>\sii$, as shown in Figs.~\ref{fig:datacollapse066_shifted}(c) and \ref{fig:datacollapse066_shifted}(d).
In all of these cases, we observe that the slope of the logarithmic drift increases with distance to the fixed-point collision.
 More precisely, the logarithmic fits performed in 
these figures are consistent with our expectation that the slope of the drift increases linearly with $\absolute{s - \sii}$.
Our finding is a strong hint towards pseudocritical scaling,
but it is fair to note that, even with access to temperature scales that cover several orders of magnitude, it is difficult to unambiguously identify a logarithm and distinguish it, \eg, from a power law with a very small exponent.
The purpose of these fits is to verify the overall scales for the change of slope with $s$, which is captured quite well for most of the shown cases, whereas deviations from the ideal pseudocritical behavior can have different reasons.
At this point, we also want to mention that our estimator \eqref{eq:appnu} for the drifting exponent might become problematic at a strong first-order transition, because then $\xi_i/\beta$ already diverges at $\alpha_z = \alpha_{xy}$, but it is well justified within the pseudocritical regime, for which this estimator is also used in other studies. This might be one of the reasons why we do not observe convergence to $1/\nu_{\lii \perp}=s$, as one would expect from our discussion in Sec.~\ref{Sec:FirstOrderScaling}. 

\begin{figure}[t]
\includegraphics[width=\linewidth]{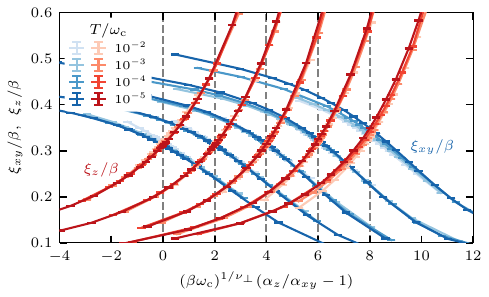}
\caption{%
Scaling collapse of the normalized correlation lengths $\xi_{xy}/\beta$ (blue) and $\xi_z/\beta$ (red) at $s=0.64$. Results are shown for different couplings $\alpha_{xy}\in\{0.15, 0.2, 0.25, 0.317, 0.4\}$ and corresponding exponents
$1/\nu_{\perp} \in\{0.33, 0.355, 0.377, 0.42, 0.48\}$. Effective critical exponents are estimated from Fig.~\ref{fig:nudrift_below}(a) and we use the same ones for $\xi_{xy}/\beta$ and $\xi_{z}/\beta$.
 Dashed lines indicate the shift of the origin from left to right.
}
\label{fig:nudrift_collapse}
\end{figure}
Finally, Fig.~\ref{fig:nudrift_collapse} depicts how the drift of the exponents affects the quality of the scaling collapse at fixed $s=0.64$ and for different couplings $\alpha_{xy}$. For $\xi_z/\beta$, we obtain good overlap for $\alpha_{xy}\in\{0.15, 0.2, 0.25\}$, but deviations start to increase for $\alpha_{xy}\gtrsim 0.317$, in agreement with Fig.~\ref{fig:nudrift_below}(a). We use the same exponents for the data collapse of $\xi_{xy}/\beta$,
which again is reasonably good for $\alpha_{xy} \lesssim 0.25$, but deviates more strongly for larger couplings.
All in all, Fig.~\ref{fig:nudrift_collapse} confirms all our findings from Fig.~\ref{fig:nudrift_below}.

\section{Consequences for the fully anisotropic spin-boson model\label{Sec:Ani}}

So far, we have focused on the U(1)-symmetric spin-boson model with
$\alpha_x = \alpha_y \equiv \alpha_{xy}$ and only tuned the anisotropy $\alpha_z$ along the spin-$z$ direction. In the following, we want to
discuss how our preceding findings determine the critical behavior of
the fully anisotropic model with $\alpha_x \neq \alpha_y \neq \alpha_z$.

As described in Sec.~\ref{Sec:FPstructure}, the perturbative RG predicts that all fixed points in the high-symmetry parameter manifolds are unstable towards anisotropies \cite{PhysRevB.66.024426, PhysRevB.66.024427} and, therefore, the only nontrivial fixed points, beyond the free-spin fixed point $\f$
at $\vec{\alpha}_\f =(0,0,0)$, are the strong-coupling fixed points $\lx$, $\ly$, and $\lz$. For $s<1$, the spin is always in the localized phase along the strongest dissipation strength;
without loss of generality, we assume $\alpha_x > \alpha_y > \alpha_z$.
If we tune the ratio of the two largest couplings through $\alpha_x /  \alpha_y = 1$, we can drive a transition between the two localized phases $\lx$ and $\ly$, for which the critical point is determined by the stable fixed points of the two-bath spin-boson model. In analogy to our discussion in Sec.~\ref{Sec:QC_anisotropy}, this symmetry-enhanced transition is continuous for $\si < s < 1$ and $\alpha_z < \alpha_x = \alpha_y < \alpha_{xy}^\mathrm{c}$ and first order otherwise. Therefore,
it has been important to determine the phase diagrams at finite anisotropies in Sec.~\ref{Sec:PhaseDiagAni}.
 Because of the fixed-point annihilation in the two-bath spin-boson model, there exists an extended regime at $s \lesssim \si$ and $\alpha_x = \alpha_y \ll \alpha^\ast_1$,
for which the local moment at the transition becomes extremely small. As a result, the transition between the two localized phases can become weakly first order, without the need to fine tune the parameters of the model.
Moreover, pseudocritical scaling can be observed on both sides of the fixed-point collision, in the same way as discussed in Sec.~\ref{Sec:Pseudocriticality}.

All in all, the collision of two intermediate-coupling fixed points within the high-symmetry
parameter manifolds has direct consequences for the critical behavior of the fully anisotropic spin-boson model, 
even if one cannot scan the model within its high-symmetry parameter manifolds.

\section{Conclusions and outlook\label{Sec:Con}}

In this paper, we have explored how fixed-point annihilation within a high-symmetry manifold affects
the phase transitions across this manifold tuned by anisotropy.
Using large-scale QMC simulations for a $(0+1)$-dimensional spin-boson model, we were able to study
this problem in an unprecedented manner covering temperature ranges of at least three orders of magnitude. We have uncovered an order-to-order transition between different localized phases that can be tuned from first order to second order via the bath exponent $s$. While the first-order transition is described by a discontinuity fixed point at which the inverse correlation-length exponent becomes $1/\nu_{\lii \perp} =s$, the continuous transition is determined by a critical fixed point at which the local moment is zero. 
Thereby, the anisotropic spin-boson model exhibits the exotic scenarios of both symmetry-enhanced first-order transitions as well as continuous order-to-order transitions beyond the Landau paradigm. In addition,
fixed-point annihilation provides us with a generic mechanism to tune our system towards a weak first-order transition with pseudocritical exponents that experience a logarithmically slow drift in temperature. We presented direct numerical evidence for this pseudocritical scaling, which does not only occur right after the fixed-point collision but also in-between the two intermediate-coupling fixed points before they collide. Our work unravels a rich phenomenology of quantum criticality across fixed-point collisions
representing a generic setup that is relevant for physical applications far beyond the spin-boson model.

One motivation to investigate the fixed-point collision in the anisotropic spin-boson model is to get a better 
understanding of the critical properties of nonlinear sigma models with a topological $\theta$ term which potentially exhibit such an RG scenario in the context of deconfined criticality
\cite{PhysRevX.5.041048, PhysRevX.7.031051, PhysRevB.102.201116, PhysRevB.102.020407}. It has been pointed out  that the spin-boson model is a $(0+1)$-dimensional representation of such a Wess-Zumino-Witten model \cite{PhysRevB.102.201116, PhysRevB.106.L081109}; in our case, the competition between the spin-Berry phase and the retarded spin interaction leads to a complex fixed-point structure. In the spin-boson model, the order parameter is an SO(3) vector which exhibits fixed-point annihilation
within its high-symmetry manifold and we consider an anisotropy along one of its components to tune through the different transitions. To explain the nontrivial scaling behavior in two-dimensional SU(2) quantum magnets, it has been conjectured that a fixed-point annihilation occurs in a $(2+\epsilon)$-dimensional Wess-Zumino-Witten model with $\epsilon \lesssim 1$ such that the transition at $\epsilon = 1$ is of weak first order due to its proximity to the fixed-point collision \cite{PhysRevB.102.201116, PhysRevB.102.020407, footnoteXXZspinboson}.
The SO(5) theory of this transition combines the antiferromagnetic and valence-bond-solid order parameters such that the transition between the two orders can be tuned by an anisotropy similar to our scenario. It is important to note that the SO(5) symmetry of the deconfined transition emerges from the corresponding microscopic Hamiltonians \cite{PhysRevX.5.041048}, whereas in the spin-boson model the SO(3) symmetry is manifest at the transition. Although this might lead to differences in the mechanisms that drive the transitions, the phenomenology of these phase transitions is determined by their fixed-point structure.
In both scenarios, numerical simulations reveal drifting critical exponents
\cite{PhysRevLett.104.177201, PhysRevB.88.220408, Jiang_2008, PhysRevLett.110.185701, PhysRevLett.101.050405, doi:10.1126/science.aad5007, PhysRevX.5.041048}
 as well as symmetry enhancement \cite{Zhao:2019aa, PhysRevB.99.195110} at the weak first-order transition. Therefore, spin-boson models provide a valuable platform to study this phenomenology using large-scale numerical simulations, as the reduced dimensionality is an advantage at reaching large ranges of system sizes which are not as easily accessible in higher-dimensional models. In future, it will be interesting to investigate whether the spin-boson model can be modified in such a way that additional interactions lead to an emergent (and not built-in) symmetry at criticality, as suggested in Ref.~\cite{PhysRevB.106.L081109}, or even to an approximate emergent symmetry.
While in deconfined criticality fractionalized excitations play an important role at driving the continuous transition, it is not clear if such excitations exist for the spin-boson model. After completion of this work, it has been pointed out that fixed-point annihilation as a mechanism for a continuous order-to-order transition does not require fractionalization \cite{Moser_2025}.
Fixed-point annihilation has also been proposed to occur in a $(1+1)$-dimensional Wess-Zumino-Witten model with long-range retarded interactions induced by dissipation \cite{martin2023stable}, for which anisotropy effects might lead to a similar phenomenology as discussed in our paper.

Spin-boson models are often embedded in Bose-Fermi Kondo models in which the spin experiences an additional Kondo coupling to a fermionic bath. Such models occur in the self-consistent solution of extended dynamical mean-field theory \cite{Si:2001aa, PhysRevB.68.115103}, but by themselves already exhibit interesting criticality.
Early analytical work has discussed the weak-coupling fixed points of the spin-boson model in the context of the Bose-Fermi Kondo model \cite{PhysRevB.66.024426, PhysRevB.66.024427}. Recently, the rich fixed-point structure of this model has been studied using QMC simulations \cite{PhysRevB.100.014439}; the occurrence of two sequential fixed-point annihilations has been suggested in the limit of spin $S\to\infty$ using an analytical approach \cite{2022arXiv220708744H}.
In our present paper, we have provided a detailed numerical study of the sequential fixed-point annihilation in the anisotropic spin-boson model. We expect the same phenomenology to apply to the Bose-Fermi Kondo model, but the computational effort to study this model is higher because the fermionic bath requires the evaluation of determinants in QMC simulations. While the precise properties of the Kondo-destruction fixed points are different from the ones considered in our work, it will be interesting to study the criticality and pseudocriticality near the two fixed-point collisions. It is an open question as to how pseudocriticality in Bose-Fermi Kondo models could affect self-consistent solutions of dynamical mean-field theory for non-Fermi liquids \cite{RevModPhys.94.035004}.
Fixed-point annihilation also occurs in purely fermionic impurity models like the underscreened Kondo model \cite{PhysRevB.57.14254} or the power-law Kondo model \cite{PhysRevB.88.195119}. This makes quantum impurity models natural candidates to explore (sequential) fixed-point annihilations and their (pseudo)critical properties.

Our work also underlines the importance of combining analytical and numerical approaches to understand complex critical behavior. Our QMC method provides exact numerical results far beyond the range of applicability of the perturbative RG, but without any knowledge of the fixed-point structure it is impossible to distinguish between the first- and second-order transitions in the pseudocritical regime. In particular, the fact that we find pseudocritical exponents close to the fixed-point collision, which hardly show any drift over several orders of magnitude in temperature but seem to depend on the initial couplings, can easily lead to wrong conclusions. This is even true for the continuous transition if we tune the anisotropy through the regime right in-between the two intermediate-coupling fixed points. Although the critical fixed point eventually determines the true critical exponent at a fixed bath exponent $s$, the slow drift of the effective critical exponent could easily be misinterpreted if we did not know the precise fixed-point couplings. Knowledge of the detailed fixed-point structure from analytical and numerical considerations is rarely available for more complicated models, making the spin-boson model an ideal platform to study the consequences of intricate RG phenomena like fixed-point annihilation.

All in all, fixed-point annihilation is not just an abstract concept but applies to quantum systems that are as simple as a spin coupled to its environment. The competition of two bath components is already enough to find signatures of this RG phenomenon in the critical properties of the anisotropic spin-boson model. The fact that fixed-point annihilation has been proposed in a variety of contexts in different areas of physics suggests that our numerical findings are relevant far beyond the spin-boson model. It will be interesting to investigate further analogies in the future.

\begin{acknowledgments}
I thank Matthias Vojta for our previous collaboration on Ref.~\cite{PhysRevLett.130.186701}, from which I learned a lot about the spin-boson model, and for ongoing discussions. I also thank Lukas Janssen for helpful discussions 
and for pointing out Refs.~\cite{PhysRevLett.35.477, PhysRevB.26.2507} on discontinuity fixed points.
Moreover, I acknowledge discussions with David Moser, Michael Scherer, Yuan Wan, and Zhenjiu Wang.
This work was supported by the Deutsche Forschungsgemeinschaft through the W\"urzburg-Dresden Cluster of Excellence on Complexity and Topology in Quantum Matter -- \textit{ct.qmat} (EXC 2147, Project No.\ 390858490).
The authors gratefully acknowledge the computing time made available to them on the high-performance computer at the NHR Center of TU Dresden. This center is jointly supported by the Federal Ministry of Education and Research and the state governments participating in the NHR \footnote{\url{https://www.nhr-verein.de/unsere-partner}}.
I also acknowledge support from the Max Planck Institute for the Physics of Complex Systems, where I have performed the first part of the numerical simulations.
\end{acknowledgments}

\appendix

\section{Renormalization-group flow close to the fixed-point collision}
\label{App:RGquad}

To get a better understanding of the slow RG flow close to the fixed-point annihilation,
we consider the minimal beta function of Eq.~\eqref{eq:RGquadratic},
\begin{align}
\label{eq:betaf}
\beta(\bar\alpha) \equiv
\frac{d\bar{\alpha}}{d\ln \mu} = a \left(s - s^\ast \right) - b \left( \bar{\alpha} - \bar{\alpha}^\ast \right)^2 \, ,
\end{align}
which is an approximation to the real beta function that becomes valid near the fixed-point collision.
For $s>s^\ast$, Eq.~\eqref{eq:betaf} contains two intermediate-coupling fixed points at $\bar{\alpha}_\pm = \bar{\alpha}^\ast \pm \sqrt{a\left(s-s^\ast\right)/b}$.
If we tune the external parameter $s$ to $s^\ast$, the two fixed points collide at $\bar{\alpha} = \bar{\alpha}^\ast$ and disappear into the complex plane for $s < s^\ast$.
In the following, we explicitly solve the RG equation for $s \gtrsim s^\ast$ and $s \lesssim s^\ast$.
To this end, we apply separation of variables to Eq.~\eqref{eq:betaf} and solve
\begin{align}
\label{eq:betaint}
\ln(\mu / \mu_0) = \int_{\bar{\alpha}_0}^{\bar{\alpha}} \frac{d \bar{\alpha}^\prime}{a \left(s - s^\ast \right) - b \left( \bar{\alpha}^\prime - \bar{\alpha}^\ast \right)^2}
\end{align}
before and after the fixed-point collision.

\subsection{After the fixed-point collision ($s < s^\ast$)}

To solve the integral in Eq.~\eqref{eq:betaint},
we substitute $x' = \sqrt{b/(a \absolute{s-s^\ast})} \left( \bar{\alpha}' - \bar{\alpha}^\ast \right)$ and obtain
\begin{align}
\nonumber
\ln(\mu / \mu_0) 
	&= 
	- \frac{1}{\sqrt{ba \absolute{s - s^\ast}}} \int_{x_0}^{x} \frac{dx'}{1+x'^2}
	\\
	&=
	- \frac{1}{\sqrt{ba \absolute{s - s^\ast}}} \, \arctan(x') \big\vert_{x_0}^{x} \, .
\label{eq:ln1}
\end{align}
Note that $\arctan(x')$ only changes rapidly around $x'=0$ and approaches $\pm \pi/2$ for $\absolute{x'} \gg 1$.
If we start our RG flow at $x_0 = 0$ and only evolve to $x\ll 1$, we can expand $\arctan(x)$ such that the RG flow of our original coupling parameter becomes
\begin{align}
\label{eq:apppseudo}
\bar{\alpha}(\mu) = \bar{\alpha}^\ast -  a \absolute{s - s^\ast} \ln(\mu / \mu_0) \, ,
\end{align}
as also stated in Eq.~\eqref{eq:RGsc2}. Note that this approximation is only valid for
$\absolute{\bar\alpha - \bar{\alpha}^\ast} \ll \sqrt{a \absolute{s-s^\ast}/b}$, for which the beta function can be considered constant. In this small
parameter regime, our system exhibits an extremely slow RG flow
that leads to pseudocritical scaling behavior, as discussed in Sec.~\ref{Sec:Pseudocriticality},
and affects the flow towards strong couplings for all $\bar{\alpha} < \bar{\alpha}^\ast$.
In particular, if we start our RG flow at $x_0 \ll -1$ and end at $x \gg 1$, \ie, our initial and final couplings fulfill
$\absolute{\bar{\alpha}^\prime - \bar{\alpha}^\ast} \gg \sqrt{a \absolute{s - s^\ast} / b}$,
we arrive at the well-known result
\begin{align}
\label{eq:quasiuni}
\frac{\mu}{\mu_0} = \exp\left(- \frac{\pi}{\sqrt{ba\absolute{s-s^\ast}}}\right) \, .
\end{align}
For $s \to s^\ast$, the infrared scale $\mu$ gets exponentially suppressed.
In particular, the total RG time $t=\absolute{\ln (\mu/\mu_0)}$ does not depend on the initial conditions and is only
determined by a small interval around $\bar{\alpha}^\ast$. As a result, fixed-point annihilation provides a generic mechanism
to generate an extremely slow RG flow over a wide range of parameters $\bar{\alpha} \ll \bar{\alpha}^\ast$ and $s \lesssim s^\ast$,
which leads to a separation of hierarchies and quasiuniversality.

\subsection{Before the fixed-point collision ($s > s^\ast$)}

In the presence of the two intermediate-coupling fixed points, we use the same substitution as before to obtain
\begin{align}
\nonumber
\ln(\mu / \mu_0) 
	&= 
	\frac{1}{\sqrt{ba \absolute{s - s^\ast}}} \int_{x_0}^{x} \frac{dx'}{1- x'^2}
	\\
	&=
	\frac{1}{\sqrt{ba \absolute{s - s^\ast}}} \, \artanh(x') \big\vert_{x_0}^{x} \, .
\end{align}
Here, we only consider the RG flow for starting points $x_0$ between the two fixed points at $x_\pm = \pm 1$.
For simplicity, we assume that we start the RG flow at $x_0 = 0$. Because the RG time diverges when approaching the stable fixed point at $x_- = -1$,
we parametrize $x= x_- +\Delta x$. Using $\artanh(x) = \frac{1}{2} \ln\big(\frac{1+x}{1-x}\big)$,
we obtain
\begin{align}
\label{eq:RGflow2}
\frac{\mu}{\mu_0} = \left(\frac{\Delta x}{2 -\Delta x}\right)^{\nu} \, .
\end{align}
Here, we have introduced the inverse correlation-length exponent $1/\nu = 1/\absolute{\nu_\pm} = 2 \sqrt{b a \absolute{s-s^\ast}}$ at the two fixed points.
Note that our results can be transformed back to our original variables via the relation 
$\Delta x = 2 \, \Delta \bar{\alpha}  / \absolute{\bar{\alpha}_+ - \bar{\alpha}_{-}}$, but we will keep the $x$ notation in the following, because then all couplings are normalized by the distance between the two fixed points.

For $\Delta x \to 0$, \ie, $\mu/\mu_0 \to 0$, we obtain the asymptotic behavior close to the stable fixed point,
\begin{align}
\label{eq:limit0}
\frac{\mu}{\mu_0} = \left( \frac{\Delta x}{2} \right)^\nu
\, .
\end{align}
In analogy to any isolated attractive fixed point, we obtain a power-law dependence of the infrared RG scale $\mu$ on $\Delta x$, which
is determined by the correlation-length exponent. If we tune $s\to s^\ast$, \ie, towards the fixed-point collision, $\nu \to \infty$ leads to
a very slow RG flow. This RG flow is qualitatively different from the pseudocritical behavior of Eq.~\eqref{eq:apppseudo} experienced for $s<s^\ast$.
 Note that the same power law can be observed if we approach the stable fixed point from $x < x_{-}$. Moreover, the same scaling behavior holds near the unstable fixed point $x_+$, but the direction of the RG flow gets inverted.

We also study the RG flow across the maximum of the beta function at $\bar{\alpha}^\ast$. Again, we choose $x_0=0$ but now we parametrize $x=- \Delta \tilde{x}$ with $0 < \Delta\tilde{x} \ll 1$.
Then, we can approximate $\artanh(-\Delta \tilde{x}) = -\Delta\tilde{x} + \mathcal{O}(\Delta\tilde{x}^3)$ and obtain
\begin{align}
\label{eq:quasi2}
\frac{\mu}{\mu_0}
	=
	\exp( - 2 \nu \, \Delta\tilde{x} ) \, .
\end{align}
Equation \eqref{eq:quasi2} has the same functional form as Eq.~\eqref{eq:apppseudo} and, in particular, the same dependence on $\absolute{s-s^\ast}$. This becomes apparent if we insert $2\nu = 1/\sqrt{b a \absolute{s-s^\ast}}$ and transform back to our original variables; then even prefactors are the same as in Eq.~\eqref{eq:apppseudo}.
As a result, pseudocritical scaling also occurs for $s > s^\ast$ right between the two intermediate-coupling fixed points. However, the couplings for which this phenomenon occurs are restricted to the parameter range between the two fixed points and therefore become increasingly narrow close to the fixed-point collision.

\begin{figure}[t]
\includegraphics[width=\linewidth]{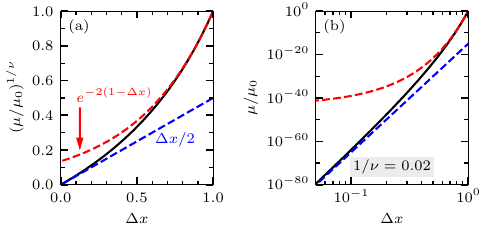}
\caption{%
Reduction of the RG scale $\mu$ as a function of the distance $\Delta x$ to the stable fixed point at $x_{-}=-1$, according to the exact solution in Eq.~\eqref{eq:RGflow2}.
The dashed lines indicate the asymptotic behavior for $\Delta x \to 0$ (blue) and for $\Delta x \to 1$ (red), as given by Eqs.~\eqref{eq:limit0} and \eqref{eq:quasi2}, respectively. Panel (a) illustrates the general solution for $(\mu/\mu_0)^{1/\nu}$ on a linear scale, which is valid for any $\nu$, whereas panel (b) shows a particular solution for $1/\nu = 0.02$ on a log-log scale.
}
\label{fig:RG_ana}
\end{figure}

To test the range of validity of the two approximations \eqref{eq:limit0} and \eqref{eq:quasi2} in comparison to the exact solution \eqref{eq:RGflow2}, Fig.~\ref{fig:RG_ana}(a) shows the renormalization of the RG scale as a function of the distance $\Delta x$ to the stable fixed point at $x_{-} = -1$. Here, we show $(\mu/\mu_0)^{1/\nu}$ on the vertical axis (including the correlation-length exponent) because then our plot applies to any $\nu$. We observe that approximation \eqref{eq:limit0} is valid for $\Delta x \lesssim 0.1$, whereas Eq.~\eqref{eq:quasi2} holds for $\Delta \tilde{x} = 1- \Delta x \lesssim 0.2$.
The absolute values on the vertical axis of Fig.~\ref{fig:RG_ana}(a) apply to $\nu = 1$, for which 
there is hardly any reduction of the RG scale $\mu$. To get an idea of the absolute scales close to the fixed-point annihilation, Fig.~\ref{fig:RG_ana}(b) shows $\mu/\mu_0$ for $1/\nu = 0.02$ on a log-log scale. While the RG scale gets substantially suppressed for all $\Delta x$, this effect is strongest in the pseudocritical regime for $\Delta \tilde{x} \leq 0.2$. For example, at $\Delta \tilde{x}=0.2$ we have $\mu/\mu_0 \approx 2 \times 10^{-9}$.

\section{Flow diagrams for the perturbative renormalization-group equations}
\label{App:pRG}

Although the two-loop beta function in Eq.~\eqref{eq:beta} is only valid in the asymptotic regime of small $(1-s)$ and small $\alpha_i$,
it can be used to obtain further insight into the qualitative RG flow of the anisotropic spin-boson model. In particular, it contains
the annihilation of pairs of fixed points which occurs beyond the range of applicability of the weak-coupling perturbative RG \cite{PhysRevB.90.245130, PhysRevLett.130.186701}.
In the following, we solve Eq.~\eqref{eq:beta} as if it was an exact equation because we expect that the qualitative features
that are characteristic for the fixed-point collision remain true for the generic case. In particular, our analysis of Eq.~\eqref{eq:beta} 
allows us to tune $\alpha_x = \alpha_y \equiv \alpha_{xy}$ and $\alpha_z$ independently across the fixed-point annihilation, which is not possible with Eq.~\eqref{eq:RGquadratic}. Our discussion is based on the perturbative RG results presented in Refs.~\cite{PhysRevB.61.15152, PhysRevB.66.024426, PhysRevB.66.024427}.

\begin{figure}[tbp!]
\includegraphics[width=\linewidth]{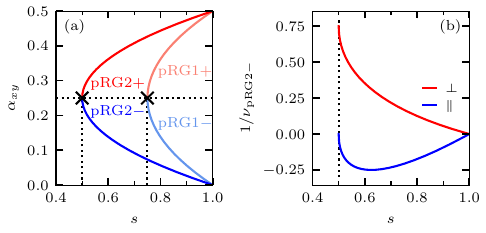}
\caption{%
(a) Fixed-point collision of $\mathrm{pRG}1/2\pm$ as a function of the bath exponent $s$ for the two-loop
beta function in Eq.~\eqref{eq:beta}. Dashed lines and crosses indicate the coordinates of the collisions.
(b) Inverse correlation-length exponents at the stable fixed point $\mathrm{pRG}2-$ in ($\parallel$) and out of ($\perp$) the SU(2)-symmetric manifold.
}
\label{fig:RG_perturbative_coll}
\end{figure}

The beta function in Eq.~\eqref{eq:beta} contains two pairs of nontrivial intermediate-coupling fixed points, \ie,
in the $xy$ plane at $\vec{\alpha}_{\mathrm{pRG}1\pm} = (\alpha_{\mathrm{pRG}1\pm}, \alpha_{\mathrm{pRG}1\pm}, 0)$
with $\alpha_{\mathrm{pRG}1\pm} = \frac{1}{4} [ 1 \pm \sqrt{1-4\left(1-s\right)} ]$ and for the SU(2)-symmetric case at 
$\vec{\alpha}_{\mathrm{pRG}2\pm} = (\alpha_{\mathrm{pRG}2\pm}, \alpha_{\mathrm{pRG}2\pm},  \alpha_{\mathrm{pRG}2\pm})$
with $\alpha_{\mathrm{pRG}2\pm} = \frac{1}{4} [ 1 \pm \sqrt{1-2\left(1-s\right)} ]$. The fixed-point collisions occur at
$s^\ast_\mathrm{pRG1} = 0.75$ and $s^\ast_\mathrm{pRG2} = 0.5$, for which $\alpha^\ast_\mathrm{pRG1/2}=0.25$,
as illustrated in Fig.~\ref{fig:RG_perturbative_coll}(a).
Note that the strong-coupling fixed points, which are beyond the range of validity of the perturbative RG, do not exhibit the correct $\alpha_{xy} \to \infty$ behavior for $s\to 1$ that is found in our QMC simulations.

We also calculate the inverse correlation-length exponents
$1/\nu_{\mathrm{pRG2} \pm \parallel} = 1-2\left(1-s\right) \pm \sqrt{1-2\left(1-s\right)}$ and
$1/\nu_{\mathrm{pRG2} \pm \perp} = 1-\frac{1}{2}(1-s) \pm \sqrt{1-2\left(1-s\right)}$
within and perpendicular to the SU(2)-symmetric manifold, respectively.
For $s\to 1$, the predictions of the perturbative RG for the fixed point $\crii$ are well controlled and we obtain
the asymptotically-exact results stated in Eqs.~\eqref{eq:exp_nupara} and \eqref{eq:exp_nu}.
Figure~\ref{fig:RG_perturbative_coll}(b) shows how the in- and out-of-plane exponents for fixed point $\mathrm{pRG2}-$ evolve towards $s \to s^\ast_{\mathrm{pRG}2}$.
We find that the in-plane exponent $1/\nu_{\mathrm{pRG2} - \parallel} \leq 0$ vanishes $\propto (s-s^\ast_\mathrm{pRG2})^{1/2}$ for $s\to s^\ast_\mathrm{pRG2}$,
which generically holds near the fixed-point collision \cite{PhysRevLett.130.186701},
whereas the out-of-plane exponent $1/\nu_{\mathrm{pRG2} - \perp} \geq 0$ remains finite for $s\to s^\ast_\mathrm{pRG2}$
and exhibits a leading correction $-\sqrt{2} \, (s-s^\ast_\mathrm{pRG2})^{1/2}$ that has the same functional dependence and the same prefactor of $(-\sqrt{2})$
as the leading term in $1/\nu_{\mathrm{pRG2} - \parallel}$.
The qualitative behavior of $1/\nu_{\mathrm{pRG2}-\parallel/\perp}$ is in good agreement with
our QMC results presented in Fig.~\ref{fig:exponents}(a). However, the strong-coupling exponents are not expected to be reliable and therefore omitted in 
Fig.~\ref{fig:RG_perturbative_coll}(b). In particular, $1/\nu_{\mathrm{pRG2} \pm \parallel}$ evolve very differently as a function of $s$.
Our QMC results for the dissipative $S=1/2$ problem, as presented in Fig.~\ref{fig:exponents}(a), reveal that $1/\nu_{\crii \parallel}$ and $1/\nu_{\qcii \parallel}$ are very close to each other \cite{PhysRevLett.130.186701}
and in the large-$S$ limit they only differ by their sign \cite{PhysRevB.106.L081109}.
On the other hand, our QMC results suggest that the out-of-plane scaling is very different at the two fixed points,
as it is also the case in the perturbative RG solution.

The two correlation-length exponents $1/\nu_{\mathrm{pRG2} - \parallel}$ and $1/\nu_{\mathrm{pRG2} - \perp}$ determine the speed of the RG flow near the fixed point $\mathrm{pRG}2-$ along different directions in parameter space. Because the evolution of the two exponents is in qualitative agreement
with our QMC results, we will analyze the RG flow of the two-loop beta function in more detail in the following.

\begin{figure}[t]
\includegraphics[width=\linewidth]{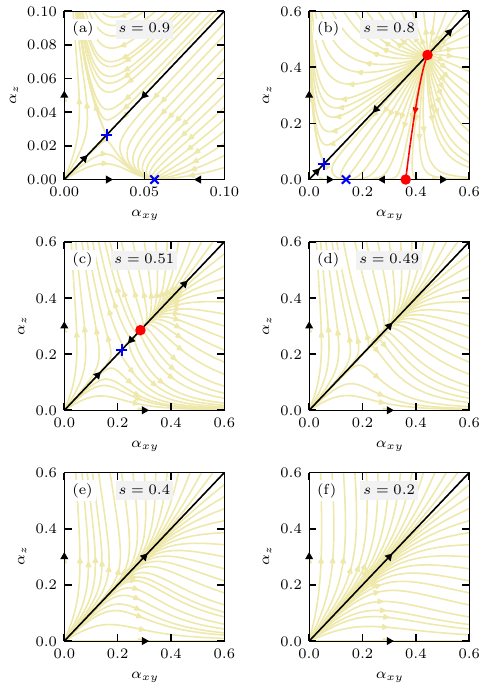}
\caption{%
RG flow diagrams for the two-loop beta function \eqref{eq:beta} as a function
of $\alpha_{xy}$ and $\alpha_z$ and for different bath exponents $s$.
The intermediate-coupling fixed points $\mathrm{pRG1/2-}$ ($\mathrm{pRG1/2+}$) are marked by
blue crosses (red circles). The red line in (b) defines the separatrix between the critical and the localized phase in the $xy$ plane.
}
\label{fig:RG_perturbative}
\end{figure}

Figure \ref{fig:RG_perturbative} shows the RG flow of the two-loop beta function \eqref{eq:beta}
for different bath exponents $s$. For $s=0.9$, the weak-coupling regime around the fixed points $\mathrm{pRG1/2}-$ depicted in
Fig.~\ref{fig:RG_perturbative}(a) is still within the range of validity of the perturbative RG and therefore
representative for the exact results discussed in our main paper.
Figure \ref{fig:RG_perturbative}(b) shows the RG flow for $s=0.8$ within a broader parameter range, which now
also includes the fixed points $\mathrm{pRG1/2}+$. Although these results exceed the range of validity of the perturbative
RG, the RG flow is qualitatively consistent with our schematic picture in Fig.~\ref{fig:RGflow}(b) for $\si < s <1$ 
[note that the slope of the separatrix between $\mathrm{pRG2}+$ and $\mathrm{pRG1}+$ is quantitatively different from the one in Fig.~\ref{fig:pds08}(a)].
For $s=0.51$ depicted in Fig.~\ref{fig:RG_perturbative}(c), the fixed points in the $xy$ plane have already annihilated each other
and we are close to the second fixed-point annihilation at $s^\ast_{\mathrm{pRG2}}=0.5$ and $\alpha^\ast_{\mathrm{pRG2}}=0.25$.
It is apparent that the RG lines near the collision point run almost perpendicular to the diagonal defined by $\alpha_z = \alpha_{xy}$,
which remains true right after the collision, as demonstrated for $s=0.49$ in Fig.~\ref{fig:RG_perturbative}(d). Only for $s \ll s^\ast_{\mathrm{pRG2}}$, as
illustrated in Figs.~\ref{fig:RG_perturbative}(e) and \ref{fig:RG_perturbative}(f) for $s=0.4$ and $0.2$,
the RG flow lines start to bend in different directions again.

The RG flow diagrams in Fig.~\ref{fig:RG_perturbative} already reveal that close to the fixed-point collision the flow perpendicular to the high-symmetry line is much faster than within, so that any attempt to probe the critical properties across this region will not probe the properties of the fixed point unless we wait for an extremely large RG time (and start the RG flow extremely close to the diagonal). Details on this behavior are discussed in Sec.~\ref{Sec:Pseudocriticality} and in Fig.~\ref{fig:RG_perturbative_st_evolution}.

\section{Details on the crossing analysis}
\label{App:getexp}

\begin{figure}[t]
\includegraphics[width=\linewidth]{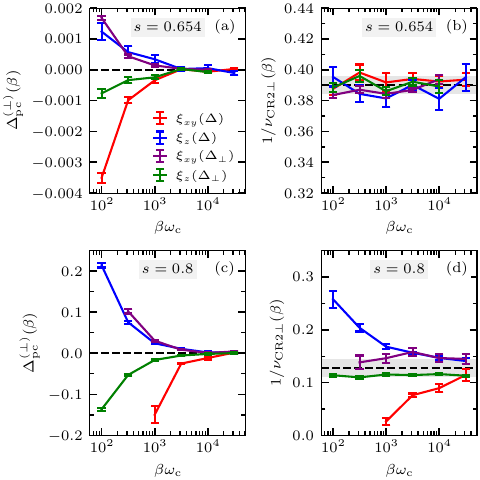}
\caption{%
Finite-size-scaling analysis for the anisotropy-driven quantum phase transition through $\crii$
for (a), (b) $s=0.654$ and (c), (d) $s=0.8$. (a), (c) Convergence of the pseudocritical couplings $\Delta^{(\perp)}_\mathrm{pc}(\beta)$, as
determined from the crossings between data sets $(\beta, r \beta)$ with $r=\sqrt{10}$, to the isotropic case indicated by the dashed line. (b), (d) Drifting exponents $1/\nu_{\crii \perp}(\beta)$ estimated at the pseudocritical couplings via
Eq.~\eqref{eq:appnu}. Dashed lines and their gray shaded background indicate our estimates $1/\nu_{\crii \perp} = 0.390(5)$ at $s=0.654$ and $1/\nu_{\crii \perp} = 0.128(15)$ at $s=0.8$.
For all data, we compare two distinct parametrizations of tuning the anisotropy $\Delta,\Delta^\perp$ through $\crii$, which we define in our main text.}
\label{fig:nu_slide}
\end{figure}

Here, we provide further information on the finite-size-scaling analysis which we use to extract the critical exponents at the continuous transition through $\crii$, as collected in Fig.~\ref{fig:exponents}.
Because the RG flow can become very slow along different directions in parameter space,
we use two distinct parametrizations to tune across the fixed point $\crii$, \ie, $\vec{\alpha} = \acrii  \left(1,1,1-\Delta\right)$
for which the anisotropy is tuned along the $z$ direction (as in the main paper) and $\vec{\alpha} = \acrii  \left(1+ \Delta^\perp,1+\Delta^\perp,1-\Delta^\perp\right)$ for which $\Delta^\perp$ is tuned perpendicular to the SU(2)-symmetric line (in the $xz$ projection). We use this comparison to convince ourselves that the slow RG flow does not affect the estimates of our critical exponents.

For both parametrizations,
Figs.~\ref{fig:nu_slide}(a) and \ref{fig:nu_slide}(c) show the temperature convergence of the pseudocritical couplings, extracted from $\xi_{i}/\beta$ via the crossings of data sets $(\beta, r \beta)$ with $r=\sqrt{10}$, for $s=0.654$ and $0.8$, respectively. For all cases, the pseudocritical couplings converge to zero, indicating that there is a single transition at the symmetry-enhanced point (and not a sequence of transitions). However, the rate of convergence can vary for different observables and parametrizations. In particular, convergence becomes substantially slower with increasing bath exponent $s$. 

Based on the crossing analysis in Figs.~\ref{fig:nu_slide}(a) and \ref{fig:nu_slide}(c), we extract the drifting inverse correlation-length exponent using Eq.~\eqref{eq:appnu}. For $s=0.654$, $1/\nu_{\crii \perp}(\beta)$ does not exhibit any visible drift in $\beta$ and all of our four estimates agree with $1/\nu_{\crii \perp} = 0.390(5)$, as shown in Fig.~\ref{fig:nu_slide}(b). For $s=0.8$, finite-size drifts are particularly strong for the two observables along the first parametrization ($\Delta$), as shown in Fig.~\ref{fig:nu_slide}(d). Nonetheless, their convergence with opposite slopes allows us to extract a reliable critical exponent of $1/\nu_{\crii \perp} = 0.128(15)$ which lies between the two data sets and agrees well with the perturbative result plotted in Fig.~\ref{fig:exponents}(a). The exponents along our second parametrization ($\Delta_\perp$) exhibit much less drifting in $\beta$ and are consistent with our estimated exponent.
We observe that for $s\to 1$ it becomes increasingly harder to extract reliable critical exponents from our numerical data, probably because additional finite-size corrections appear in the vicinity of the free fixed point.

\begin{figure}[b]
\includegraphics[width=\linewidth]{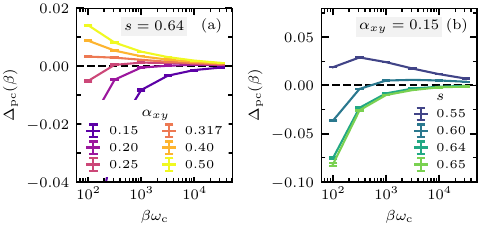}
\caption{%
Convergence of the pseudocritical couplings for the weak first-order transition (a) at fixed $s=0.64$ for different $\alpha_{xy}$ and (b) at fixed $\alpha_{xy}=0.15$ for different $s$. Results correspond to the drifting exponents in Fig.~\ref{fig:nudrift_below} and we only show the crossings obtained from $\xi_{xy}/\beta$. The dashed line indicates the isotropic case.
}
\label{fig:nu_slide_1st}
\end{figure}
Finally, we want to emphasize that the first-order transition is a single transition that occurs at $\alpha_z = \alpha_{xy}$ and that coexistence of the two ordered phases only occurs at this point. To this end, Fig.~\ref{fig:nu_slide_1st} shows the temperature convergence of the pseudocritical coupling extracted in the weak first-order regime studied in Fig.~\ref{fig:nudrift_below}. All crossings converge to the isotropic case of zero anisotropy, even for $s=0.6$ and $\alpha_{xy}=0.15$ where convergence becomes very slow. Note that Fig.~\ref{fig:nu_slide_1st} only shows the pseudocritical couplings extracted from $\xi_{xy}/\beta$; our data for $\xi_z / \beta$ show the same convergence behavior.


\begin{thebibliography}{98}
\bibitem [{foo()}]{footnoteXXZspinboson}%
  \BibitemOpen
  \href@noop {} {}\bibinfo {note} {A deconfined quantum critical point is
  supposed to exist in SU(N) generalizations of two-dimensional quantum
  magnets. Recently, it has been suggested that the weak first-order transition
  at $N=2$ could also be the result of a fixed-point collision that occurs as a
  function of $N$ \cite{doi:10.1126/sciadv.adr0634}}\BibitemShut {NoStop}%
\end{thebibliography}
\end{document}